\documentclass{aastex62}
\usepackage{epstopdf}
\begin{document}
\title{Elemental abundances distributions in (R, $V_{\phi}$) plane with LAMOST DR5 and Gaia DR2}

\author{Xilong Liang, Jingkun Zhao, Yuqin Chen, Wenbo Zuo, Jiajun Zhang, Jia Zhu, Gang Zhao}

\begin{abstract}

Since Gaia DR2 was released, many velocity structures in the disk have been revealed such as large scale ridge-like patterns in the phase space. Both kinematic information and stellar elemental abundances are needed to reveal their evolution history. We have used labels from the APOGEE survey to predict elemental abundances for a huge amount of low resolution spectra from the LAMOST survey. Deep learning with artificial neural networks can automatically draw on physically sensible features in the spectrum for their predictions. Abundances of 12 individual elements: [C/Fe], [N/Fe], [O/Fe], [Mg/Fe], [Al/Fe], [Si/Fe], [S/Fe], [Cl/Fe], [Ca/Fe], [Ti/Fe], [Mn/Fe] and [Ni/Fe] along with basic stellar labels $T_{\textrm{eff}}$, log $g$, metallicity ([M/H] and [Fe/H]) and [$\alpha$/M] for 1 063 386 stars have been estimated. Then those stars were cross matched with Gaia DR2 data to obtain kinematic parameters. We presented distributions of chemical abundances in the $V_{\phi}$ versus \textit{R} coordinate. Our results extend the chemical characterization of the ridges in the (R, $V_{\phi}$) plane to about $R = $13 kpc toward the anticenter direction. In addition, radial elemental abundance gradients for disk stars with abs(\textit{z})$ < 0.5$ kpc are investigated and we fitted a line for median abundance values of bins of stars with galactocentric distance between $R > 7.84$ kpc and $R < 15.84$ kpc. The radial metallicity gradients for disk stars are respectively -0.0475 $\pm $0.0015 for $R < 13.09$ kpc and -0.0173 $\pm $0.0028 for $R < 13.09$ kpc. Gradients for other elemental abundances are also obtained, for example, [$\alpha$/M] gradients for disk stars is 0.0030 $\pm $0.0002.

\end{abstract}


\section{INTRODUCTION}

With Gaia DR2 data release, many new kinematic sub-structures including arches and more moving groups in the solar neighbourhood were found \citep{kat18,ant18,ram18}. With $V_{\phi}$ versus \textit{R} plot, they noted that some kinematic sub-structures span a wide range of galactocentric radius. \citet{ant18} managed to explain diagonal ridges in V$_{\phi}$ vs R plane along with spiral shape in the vertical position-velocity plane by using phase mixing model. They found our galacitic disk is in an on-going vertical phase mixing state which may be induced by external satellite galaxy's perturbations. \citet{fra19} used an N-body simulation to show that the most prominent and long-lived ridges formed in the (R, $V_{\phi}$) plane could be induced by the OLR of the bar. \citet{kha19} used GALAH data and N-body simulations to show ridges could be induced by either transient perturbations (spiral arms) or interaction of stars with perturbations that are close to the plane. There are many competing and interlocking dynamical processes occurring in the Galaxy and present observation can not disentangle them yet. Chemical tagging technique is a credible way to determine where a star was born and chemical abundances information can provide lots of help for studying galactic evolution. However, the number of stars that can be observed with high resolution spectroscopy is still small. If we want to study large amount of stars in the disk, elemental abundances from low resolution spectra is a promising way.

Large Sky Area Multi-Object Fibre Spectroscopic Telescope (LAMOST) \citep{cui12,zhao12} has released a huge amount of low resolution spectra. Artificial neural networks have powerful capabilities for digging information from data. The upsurge in using artificial neural networks has already spread to Astronomy. This modern machine-learning technique has great potential for data-driven astronomy and it has already succeeded in many other fields. We chose to use a newly developed open-source python package called astroNN \citep{Leu18}\footnote{\url{ https://github.com/henrysky/astroNN}}, which has many advantages. Such as it can learn from incomplete data while take uncertainty in the training labels into account; it uses a Bayesian neural network to estimate the uncertainties of labels \citep{gal16}; it can simultaneously infer stellar basic parameters and elemental abundance labels for both high and low signal-to-noise ratio (SNR) spectra and it's efficient.

Many efforts have been taken to obtain elemental abundances from low resolution spectra. \citet{aho17} has used TheCannon to predict four labels ($T_{\textrm{eff}}$, log$g$, [Fe/H] and [$\alpha$/Fe]) from APOGEE spectra. \citet{wan19} used a method called Generative Spectrum Networks obtained those four labels for LAMOST spectra too. \citet{xia16} used kernel-based principal component analysis method obtained four basic stellar labels and carbon and nitrogen abundances from LAMOST spectra. \citet{tin17} used a neural network that consists of one hidden layer with 100 nodes with a prior based on ab initio spectral models and got 14 labels for LAMOST DR3 spectra.
By contrast, our standard deviations of residuals are smaller than theirs and our labels cover larger metallicity range than theirs. This is partly because we used much more stars than they did and partly because deep machine learning has advantage in digging information than one hidden layer neural network. We listed mean and standard deviation of residuals for both training set and test set in each residual subplots. The mean value of residuals tells us systematic difference between predictions and original values, while standard deviation of residuals tells us the accuracy of our neural network recovered labels. The two values of training set and test set are similar which means our network is stable. It has to be noted that the residuals are large for metal poor stars because metal poor stars account for relatively small percentage in the training set and most of their spectra have low SNR.

Our purpose is to study velocity structures in the thin disk by combining stellar elemental abundances and kinematics. In Section 2, we describe the data used for training our artificial neural network, detailed training processes and training results analysis. Section 3 presents our abundances results for LAMOST spectra, elemental abundances distributions in the $V_{\phi}$ versus \textit{R} plane and elemental abundances distributions along R and z direction. Finally, the main outcomes of our work are summarized in Section 4.

\section{ELEMENTAL ABUNDANCES}
\subsection{Data}

The data we used are spectra from LAMOST DR5 \citep{luo15} and labels including parameters and elemental abundances from the APOGEE DR14 \citep{abo18}. After the fifth data release, we chose to use the stellar catalogue of late A and FGK-type stars with high quality spectra (5,344,058) with measured parameters($T_{\textrm{eff}}$ , log $g$, [Fe/H], radial velocities (RV) derived by the official LAMOST Stellar Parameter Pipeline \citep{luo15}. We cross-matched the LAMOST DR5 catalogue with APOGEE DR14 catalogue and removed those stars without $T_{\textrm{eff}}$  or log $g$  values from APOGEE. As \citet{Leu18} did, only stars with surface temperature $T_{\textrm{eff}}$ between 4000K and 5500K were used to make sure the training labels are accurate \citep{gar16}. We got 32147 common stars with signal to noise ratio of g band (SNRg) from LAMOST large than 10 and named it common stellar sample. In addition, we got rid of stars with bad spectra which might caused by bad sky subtraction, bad joint between red and blue parts of spectrum, double stars, and so on. Then the training sample is obtained by re-sampling from the common stellar sample to get uniformly distributed star samples for each label to reduce the effect caused by common stellar sample's nonuniform distribution. Because neural network tends to guess values in those densest ranges to improve the accuracy. The re-sampling were taken by two steps: first we subjectively chose a reasonable range for a label and divided it into 100, 000 grids; second for each grid, its closest star was taken into the new sample. If there were more than one stars with a same smallest distance from the grid, we took the one with largest SNRg. By this way, namely over-sampling at very low density ranges through repeating real data, we constructed a generally uniformly distributed sample (training sample). However, around the marginal ranges stars may distribute too sparse, so a up limit is set that a stars can not be taken more than 10 times. After obtained uniformly distributed star samples for each label, we trained networks for each label independently. Because different labels can not be made uniformly distributed at the same time, we separately created uniform sample for each label. For each label, about 10 per cent stars were randomly chosen as test set and the rest were treated as training set. 1 063 386 stars from LAMOST DR5 catalogue with $T_{\textrm{eff}}$ between 3900K and 5600K were selected as target sample. After we obtained a trained network, it was applied to those stellar spectra to predict labels.

\subsection{Artificial Neural Network}

\subsubsection{Methodology and data treatment}

Bayesian neural network is an artificial neural network that applies Bayesian inference to obtain a posterior distribution function (PDF) on the weights of artificial neurons. Then the PDF will be used to propagate training uncertainty into predictions. Deep learning means using multi-layers to do machine learning. A technique called dropout \citep{hin12} has been widely used to avoid over-fitting. \citet{gal16} mathematically developed a theoretical framework to cast dropout training in deep neural networks as approximate Bayesian inference in deep Gaussian processes \citep{dam13}. This theory provides a tool to model uncertainty with dropout neural networks. The astroNN package runs N times Monte Carlo dropout in forward passes through the network for uncertainty estimation. It has different predictions in every forward pass through the network. It takes the mean value of predictions as the final prediction and the standard deviation of predictions as the model uncertainty. In addition to this model uncertainty, the neural network also produces a predictive uncertainty in each circulation. Predictive uncertainty captures noise inherent in the observations while model uncertainty captures our ignorance about which model generated our collected data \citep{ken17}. Finally it takes the sum of model uncertainty and predictive uncertainty of last circulation in quadrature as final predicted uncertainty to output. Only the final predicted uncertainties have been used in this paper.

First, the spectra were normalized before importing them to the neural network. We simply used the dataset module from TheCannon \citep{nes15,aho17} to do continuum normalization by gaussian smoothing. The same procedures have been applied to spectra of our target sample too. Besides the continuum normalization of the spectra, the labels were also normalized. After these steps, it is easier and faster for the neural network to converge to a minimum. The class BayesianCNNBase from astroNN was adopted to define a four layers Bayesian convolutional neural network. The first two layers are convolutional layers and the rest are dense layers with 200 and 100 neural nodes. The dropout rate was set as 0.5 and the learning process circulated 100 times passing through the neural network. All those parameters were chosen subjectively to acquire smaller bias and each label owns a neural network.

\subsubsection{Effective temperature and stellar surface gravity}

\begin{figure}[htpb]
\epsscale{.50}
\plotone{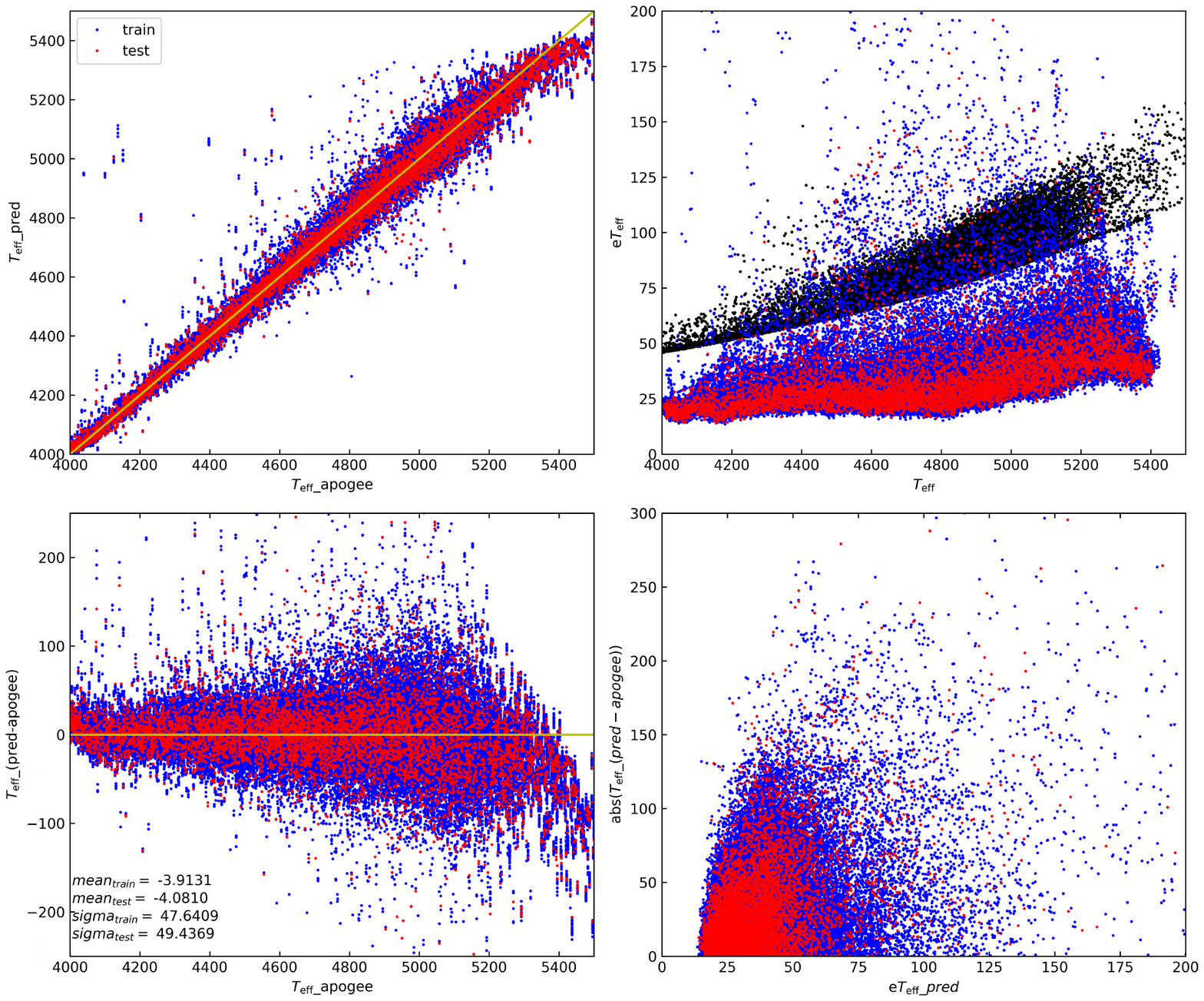}
\caption{Training and test results of effective temperature. Blue points represent training set, red points represent test set and black points in the top right subplot represent APOGEE labels. Arithmetic mean values and standard deviation values of both sets are listed in the bottom left subplot.\label{fig:f1}}
\end{figure}
\begin{figure}[htpb]
\epsscale{.50}
\plotone{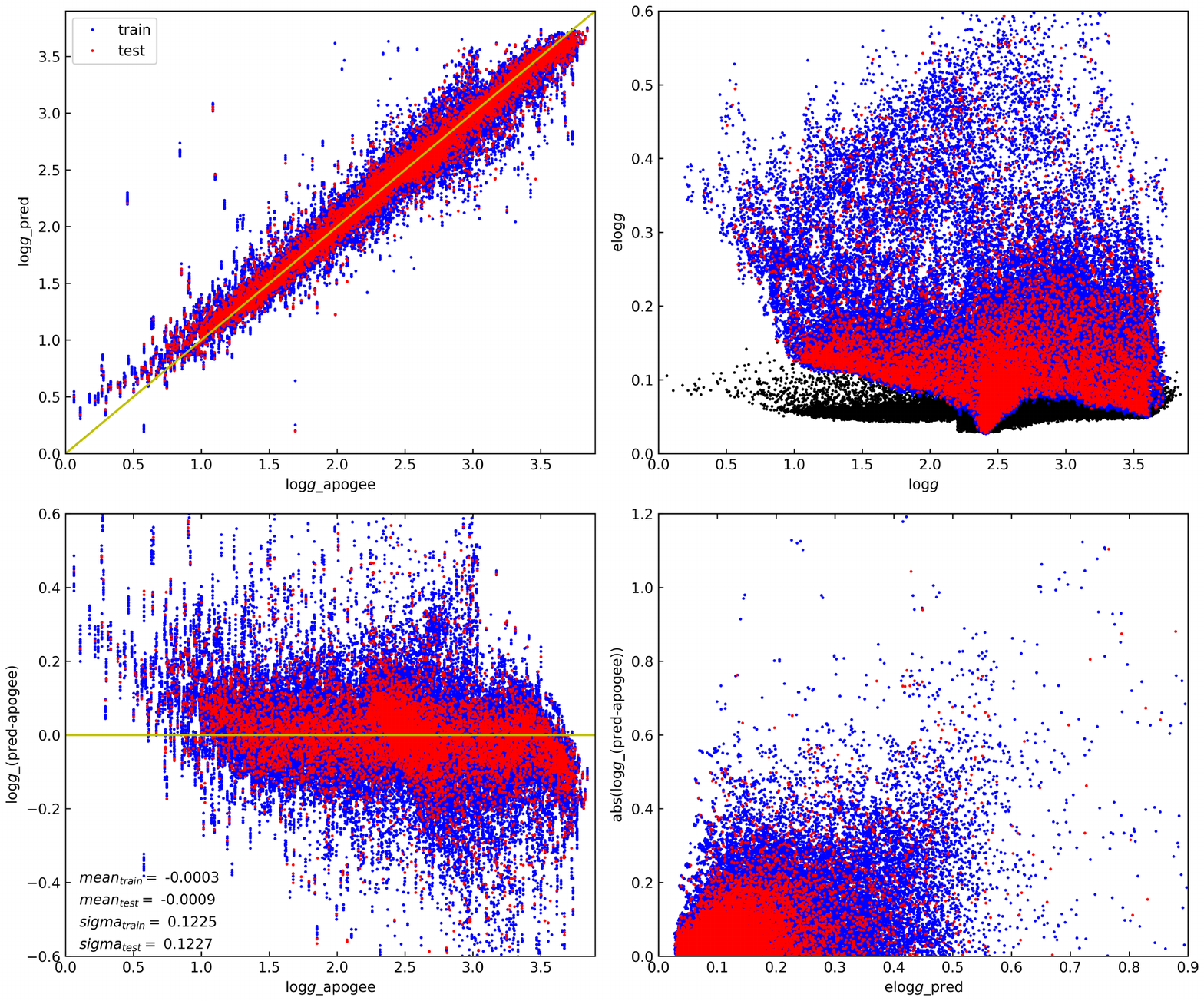}
\caption{Training and test results of surface gravity and the meaning of symbols are the same as in figure \ref{fig:f1}.\label{fig:f2}}
\end{figure}

Figure \ref{fig:f1} shows the training and test results of effective temperature. The top-left subplot shows our neural network predictions versus original APOGEE labels. The y axis is predicted effective temperature while the x axis is original APOGEE effective temperature. These blue dots represent stars of training set while these red dots represent stars of test set for all subplots in this section. The yellow line represents the diagonal line of coordinate plane. We define neural network predictions minus original APOGEE labels as residuals. Residual of effective temperature versus original APOGEE effective temperature is shown in the bottom-left subplot. The yellow line in this subplot represent $y = 0$ where the ideal expectation of residual is. We present the mean and sigma value of residuals of both training set and test set in the bottom left of this subplot. The arithmetic mean values of residuals of both training set and test set are not zero means there is systematic bias between predicted $T_{\textrm{eff}}$ and APOGEE $T_{\textrm{eff}}$ but they are only about 4 K. Sigma value is the standard deviation which is a measure of the spread of residuals and the smaller sigma is, the better prediction is. This value merely shows how well our predictions compared to the APOGEE labels and it should not be used as the uncertainty of the whole sample. Original APOGEE effective temperature themselves are not necessarily real value and each star has its own uncertainty and the more credible APOGEE label is, the better our neural network performs. The two values of training set and test set are very similar which means our network is stable and not over fitted. Through cross-validation, it shows predicted label inferred from LAMOST spectra agree well with APOGEE label. However, revealed from the residual subplot, it's not very good at the marginal parts, which is mainly due to the nonuniformity of original APOGEE labels of common stellar sample. Stars with original APOGEE effective temperature $T_{\textrm{eff}}$ larger than 5400 K have much less star numbers per unit interval than its neighbourhood and the truncation at 5500 K makes it worse. To reduce the impact of nonuniformity, we generated new uniform sample from the original common stellar sample as mentioned before. When created new $T_{\textrm{eff}}$ sample, stars within low density ranges were taken repeatedly. Even though this method can add stellar numbers, it can not add stellar diversity, thus sampling technique can only make a generally uniformly distributed sample but it can not completely eliminate the nonuniformity effect of original label distribution. As the figure shows, we recommend predicted effective temperatures within (4000, 5400) K with small uncertainties are trustable. These two subplots in the right of figure \ref{fig:f1} show uncertainty distributions. The top one shows uncertainties of effective temperatures distribute with effective temperatures. Black dots represent uncertainties of labels versus labels from APOGEE and blue and red dots respectively represent predicted uncertainties of effective temperatures versus effective temperatures uncertainties of training set and test set. It is obvious that uncertainties of APOGEE labels themselves are not horizontally uniformly distributed and the bottom bound does not start from zero. For the neural network, the sparser range a star is in, the larger predicted uncertainty is and vice versa. For those stars that are wrongly predicted, they have lager uncertainties than those stars should have been in the right places. The bottom-right subplot shows absolute values of differences between predicted labels and APOGEE labels distribute with neural network predicted uncertainties and they are positively related as can be expected. As mentioned before, the final taken uncertainty is composed of predictive uncertainty and model uncertainty, therefore the predicted uncertainty are determined by original APOGEE uncertainty and how well a stellar label being predicted and affected by random error. In bottom-right subplot, those stars with large residuals mostly have large predicted uncertainties too and the predicted uncertainty will be used to remove bad prediction. It should be noted that some stars with small residuals have large predicted uncertainty and some stars with large residuals have small predicted uncertainty and those stars may be wrongly classified when uncertainties are used to distinguish good from bad. However, these stars only take up very tiny percentage of total sample and they do not affect our statistical results.

Figure \ref{fig:f2} shows the training and test results of surface gravity. The left column of figure \ref{fig:f2} shows predicted log $g$  versus APOGEE log $g$ and residual of log $g$ versus APOGEE log $g$. The top-left cross-validation subplot shows linear relation between predicted log $g$  and APOGEE log $g$. However, the residuals are not evenly distributed along the APOGEE label which is due to nonuniformity of original APOGEE labels distribution. Sharp count density change and/or sharp uncertainty distribution change can induce this effect. Listed at the corner of bottom left subplot are arithmetic mean values and standard deviation values of residuals from training set and test set and they are $mean_{train}=-0.0003$, $mean_{test}=-0.0009$, $\sigma_{train} = 0.1225$, $\sigma_{test} = 0.1227$, respectively. The arithmetic mean values of residuals of both training set and test set are very close to zero which means there is basically no systematic bias between predicted log $g$  and APOGEE log $g$. The standard deviation values of both set are approximately the same which means there is no over fitting and for most stars the real stellar surface gravity values should be within 3 sigma from their predicted values. Those vertically adjacent points that form dashed line like features are stars with the same APOGEE labels but have different predicted labels. Since the residual are not zero, we think they are estimated deviating from APOGEE values in a similar way that are affected by random error. Though the result turns out not very bad for stars at the bottom end of log $g$, the stellar number density of common stellar sample in this region is small and we suggest those log $g$  less than 0.8 should be treated carefully when use our data. The top-right subplot shows uncertainties of log $g$  versus log $g$. The distribution of our predicted labels are determined by original distribution. Similarly, Those stars in the marginal part tend to be wrongly predicted towards the center part of log $g$ range and they have larger predicted uncertainties than those stars in the center part at first. The bottom-right subplot shows that residuals of log $g$  tend to be positively related to predicted uncertainties of log $g$.

\subsubsection{Metallicity}

\begin{figure}[htpb]
\epsscale{.50}
\plotone{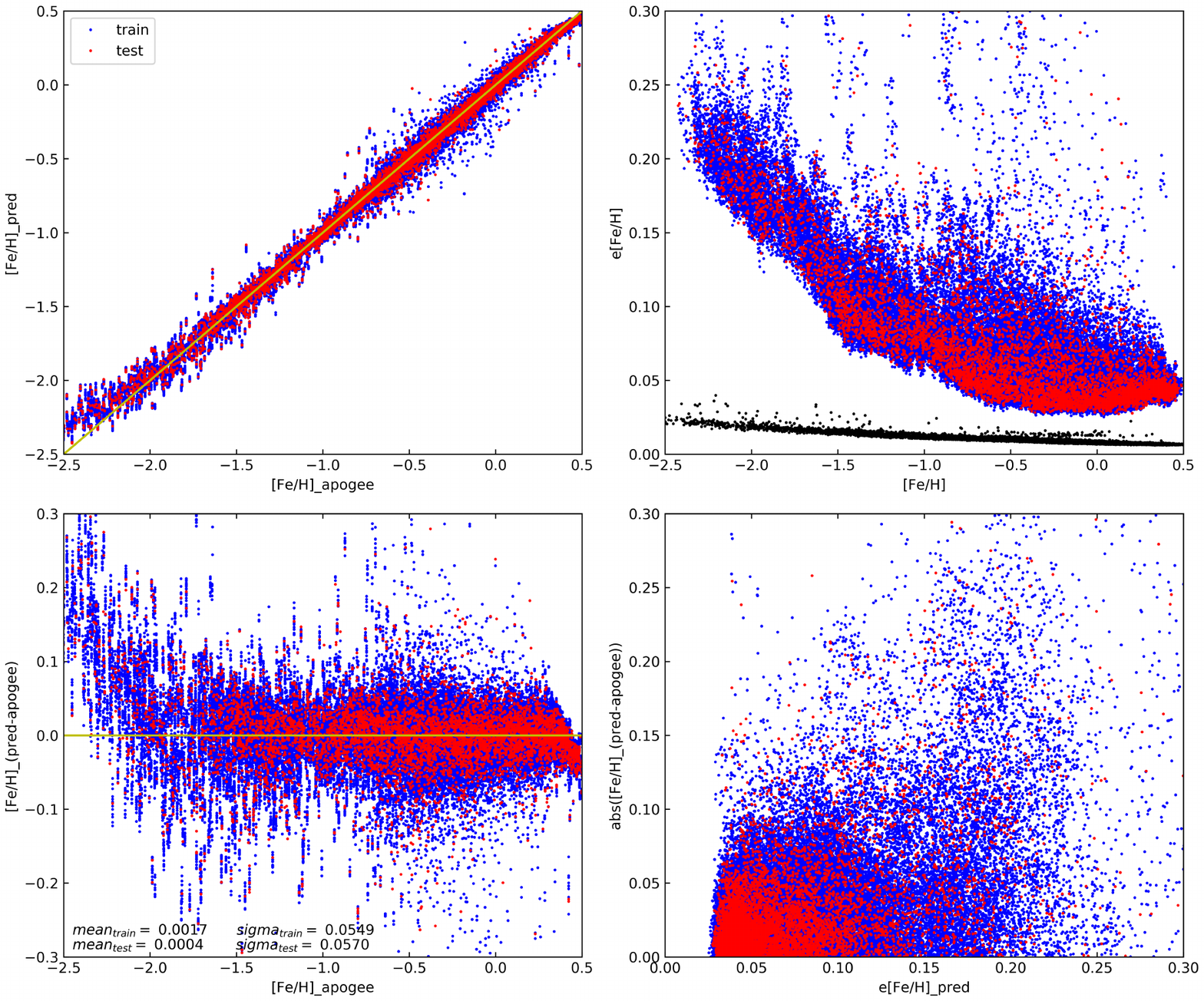}
\caption{Training and test results of [Fe/H] and the meaning of symbols are the same as in figure \ref{fig:f1}.\label{fig:feht}}
%\end{figure}
%\begin{figure}[htpb]
\plotone{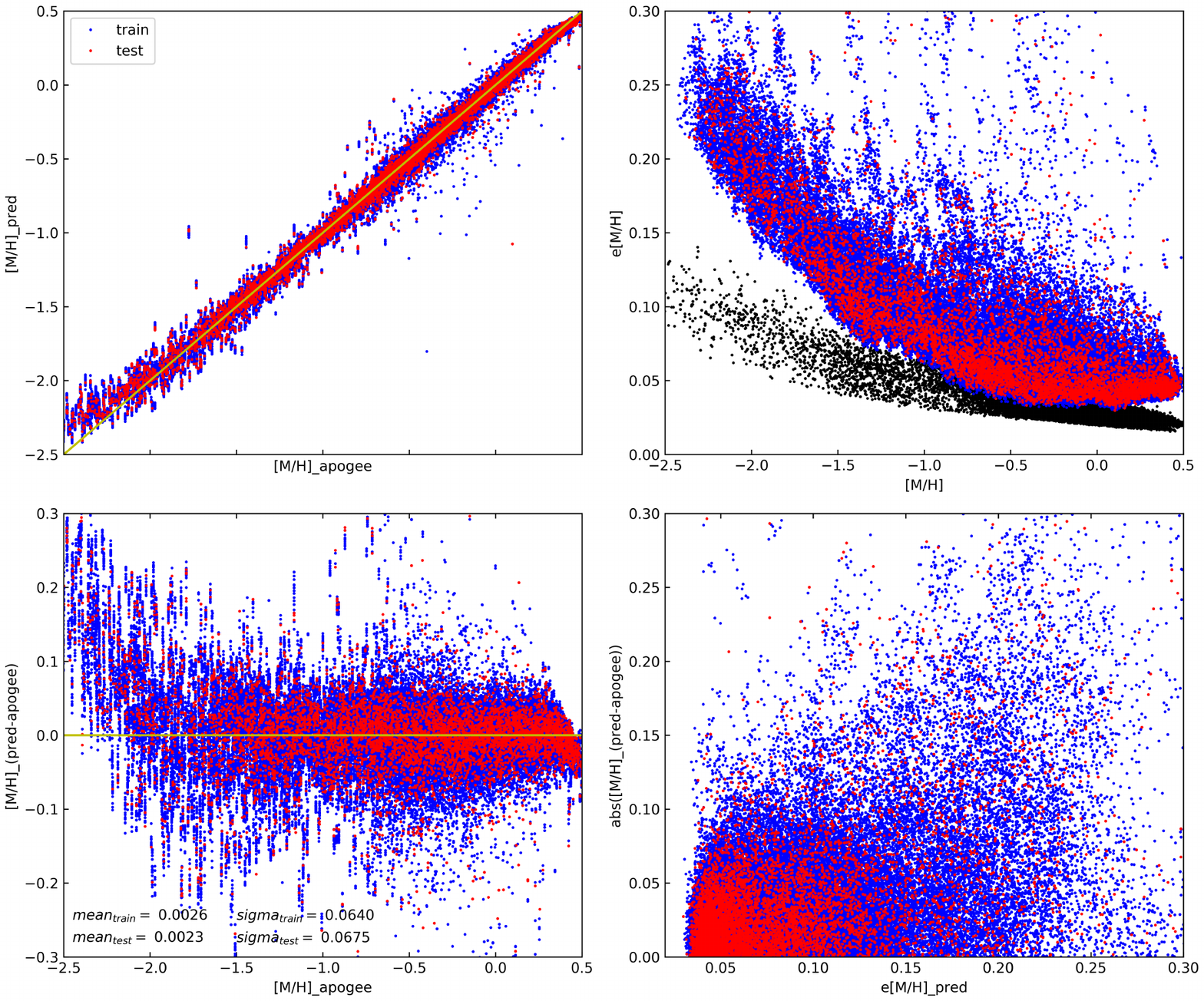}
\caption{Training and test results of [M/H] and the meaning of symbols are the same as in figure \ref{fig:f1}.\label{fig:mh}}
\end{figure}

\begin{figure}[htpb]
\epsscale{.90}
\plottwo{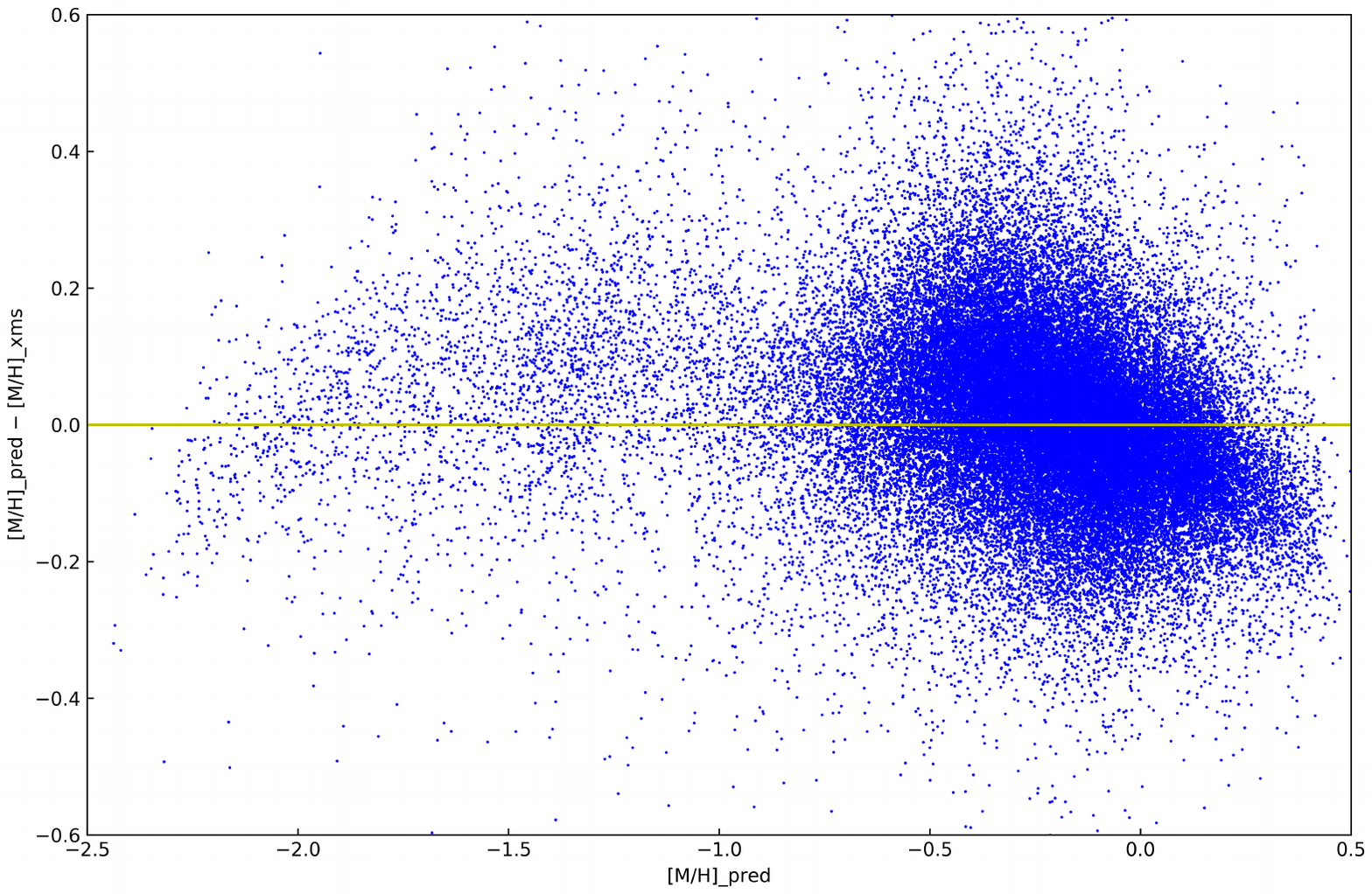}{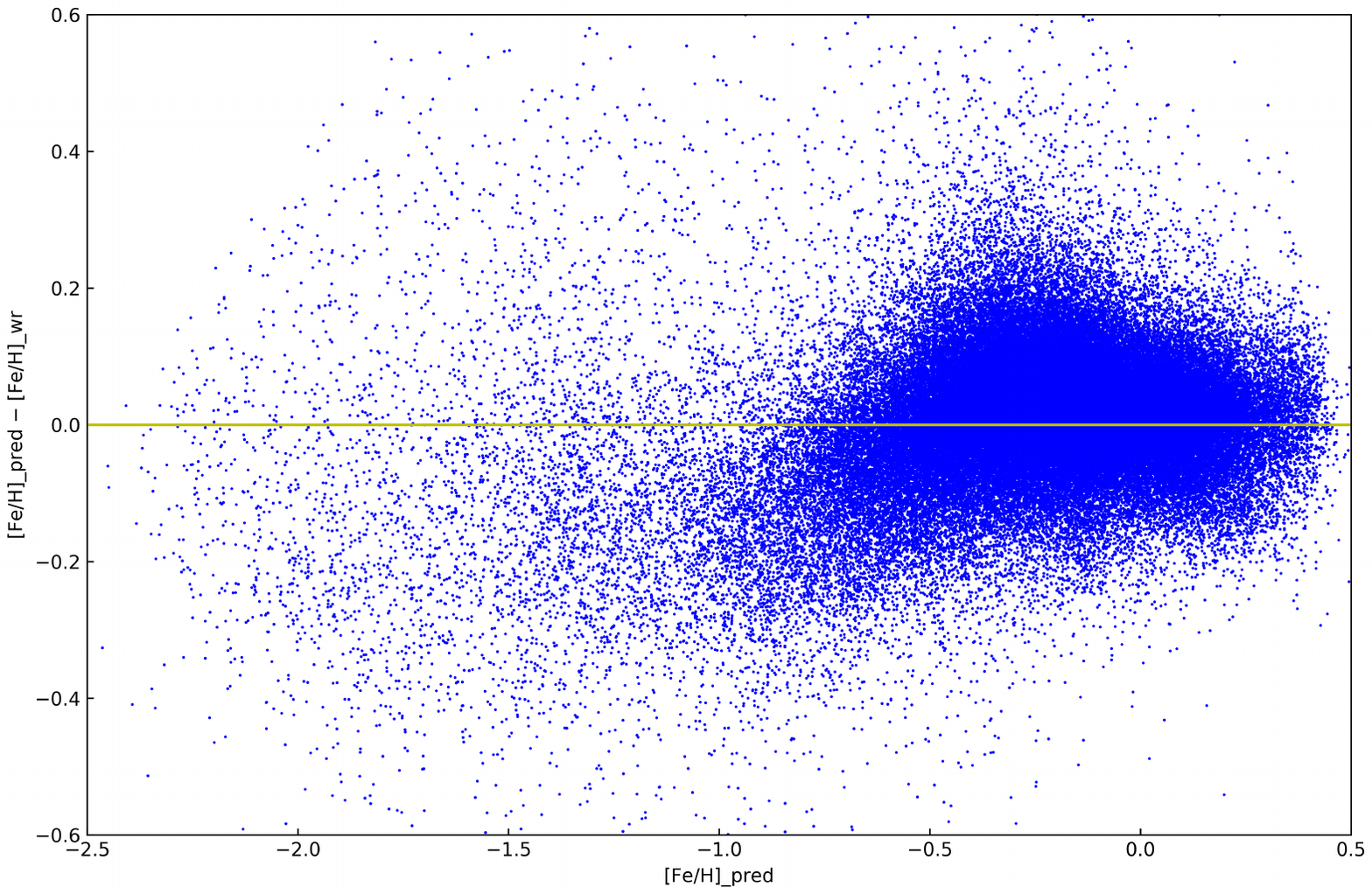}
\caption{Metallicity comparison between our predicted values with other works.\label{fig:mvs}}
\end{figure}

The APOGEE catalogue provides both [Fe/H] and [M/H], so we trained both of them respectively. Figure \ref{fig:feht} and figure \ref{fig:mh} shows the training and test results of [Fe/H] and [M/H] respectively. Figure \ref{fig:feht} and figure \ref{fig:mh} are approximately the same because iron is the dominate element for metals. The top-left cross-validation subplots show most stars are very close to the yellow diagonal line. The ideal residual should be zero. Those stars with very large positive residuals are wrongly overestimated while stars with very small negative residuals are wrongly underestimated. The mean residuals of [M/H] are 0.0026 and 0.0023 and standard deviations of residuals of [M/H] are 0.064 and 0.0675 for training set and test set respectively. The mean residuals of [Fe/H] are 0.0017 and 0.0004 and standard deviations of residuals of [Fe/H] are 0.0549 and 0.0570 for training set and test set respectively. Two top-right subplots of both [Fe/H] and [M/H] show similar distributions of our predictions though the original APOGEE uncertainties of them seem different. At both end of metallicity, the predicted values are affected by boundary effect that they tend to distribute towards denser ranges. Moreover, there are only few stars with [Fe/H]$<$-2 in the original common stellar sample. So when using our data with [Fe/H]$ <-2$, their predictions should be treated very careful.

Figure \ref{fig:mvs} shows our result compared with former researchers' work. The left subplot compares our predicted [M/H] with results of \citet{xia16} for 850,341 common stars and the right subplot compares our predicted [Fe/H] with results of \citet{wan19} for 1,058,095 common stars. Both x axes represent our predicted labels and y axes represent our predicted labels minus other researchers' result and those two yellow line represent where y$ = $0. Though all predictions are statistically in consistent with APOGEE labels, residuals of each prediction are not zero and the deviations from APOGEE labels are quite possible not towards absolute real values. Therefore differences between our predictions and their predictions of some stars can be large. For stars with our [M/H] $<$ -0.5, our predictions are systematically larger than [M/H] of \citet{xia16} while for stars with our [M/H] $>$ 0, our predictions are systematically smaller than [M/H] of \citet{xia16}. For stars with our [Fe/H] $<$ -0.5, our predictions are systematically smaller than [Fe/H] of \citet{wan19}. Those significant differences mean machine learning results can only be used to study statistical tendencies.

\begin{figure}[htpb]
\epsscale{.50}
\plotone{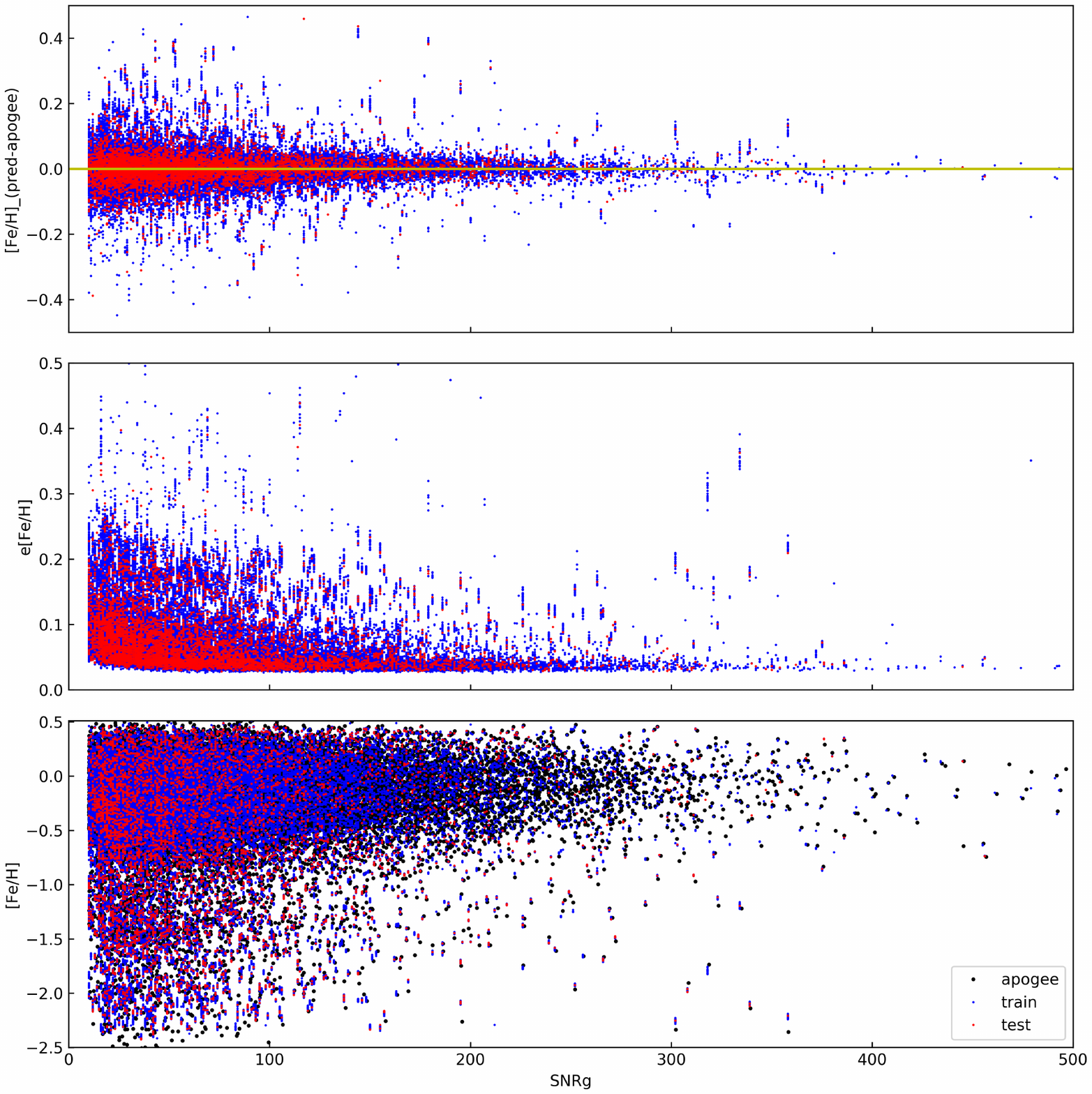}
\caption{Residuals and uncertainties of [Fe/H] and [Fe/H] themselves distribute with SNRg.\label{fig:snrfeh}}
\end{figure}

Taking metallicity as an example, figure \ref{fig:snrfeh} shows how the SNR of spectra affect predictions. The upper two subplots show residuals and neural network predicted uncertainties of training set and test set distribute with SNRg. As expected, the whole tendency is that residuals and uncertainties become smaller as SNR become larger. The lowest subplot shows [Fe/H] distributes with SNRg and metal poor stars have smaller SNR than metal rich stars on the whole. Compared to metal rich stars, metal poor stars have lower original number densities, smaller SNR and larger original uncertainties, therefore the top-right subplots in both metallicity figure haves steeper distribution of predicted uncertainties than original APOGEE uncertainties.

\begin{figure}[htpb]
\epsscale{.50}
\begin{center}
\begin{minipage}[l]{0.46\textwidth}
\leftline{\includegraphics[scale=0.44]{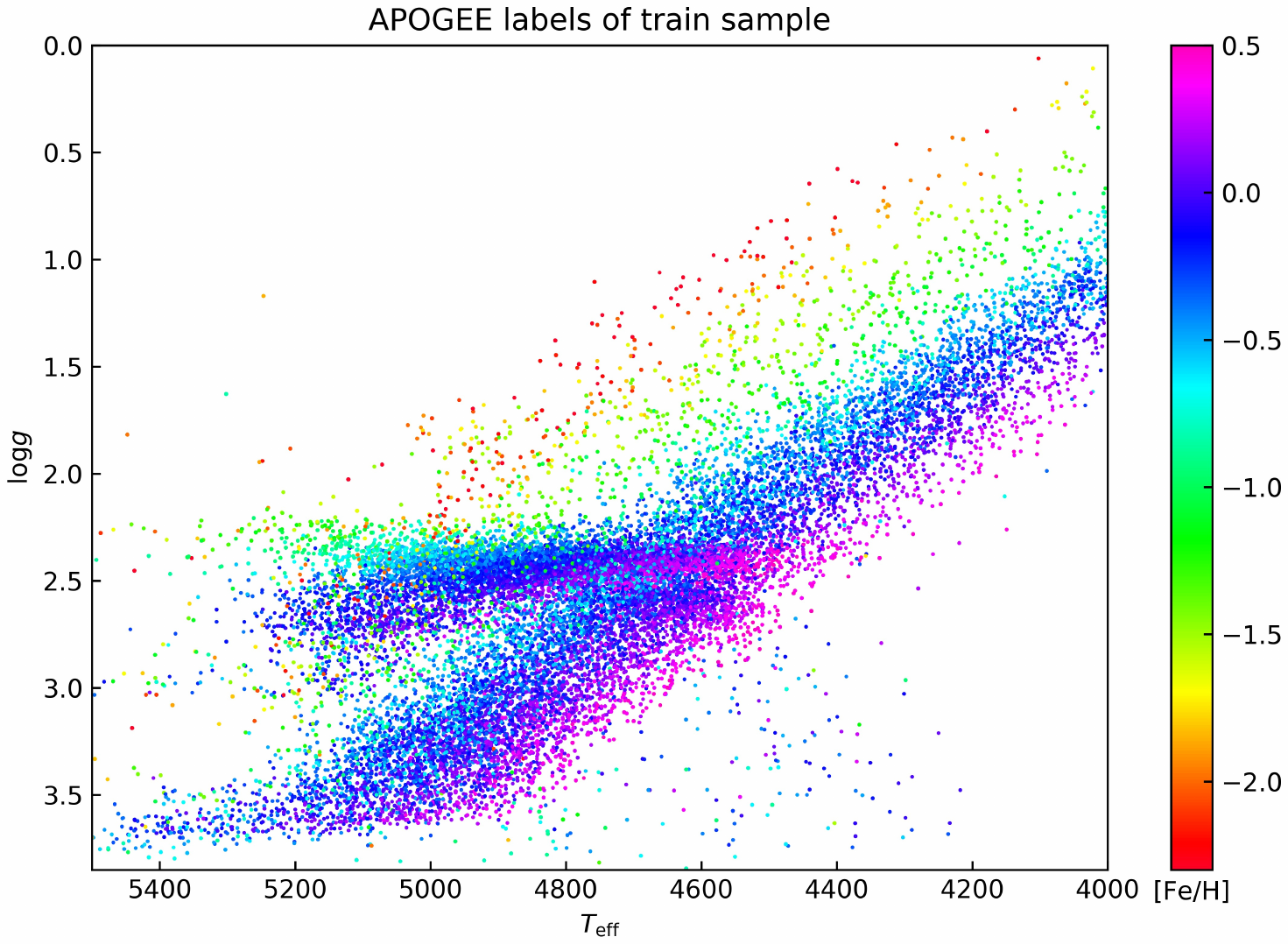}}
\centerline{(a)}
\end{minipage}
\hspace{0pt}
\begin{minipage}[l]{0.46\textwidth}
\leftline{\includegraphics[scale=0.44]{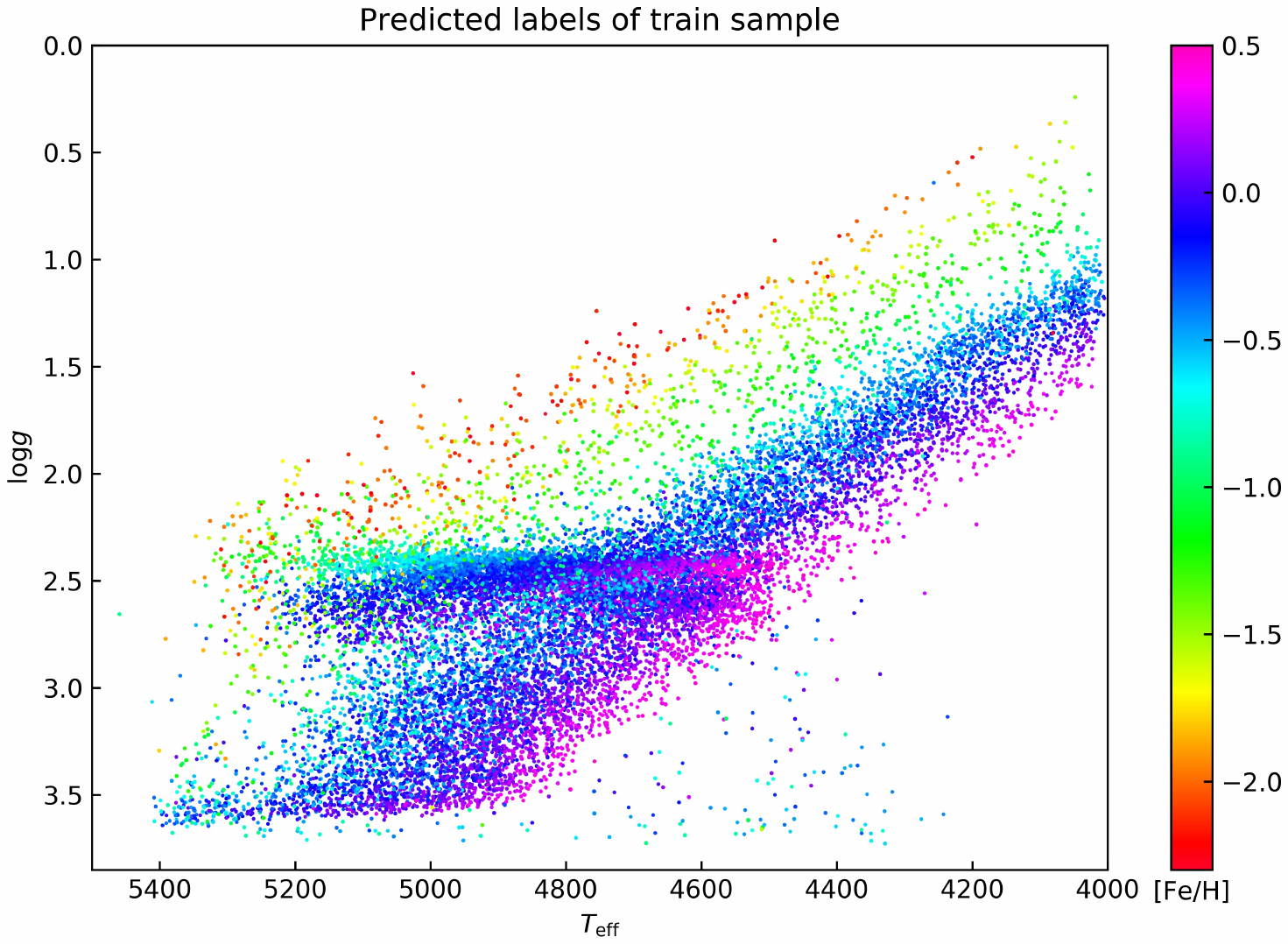}}
\centerline{(b)}
\end{minipage}
\end{center}
\vspace{0pt}
\begin{center}
\begin{minipage}[l]{0.46\textwidth}
\leftline{\includegraphics[scale=0.44]{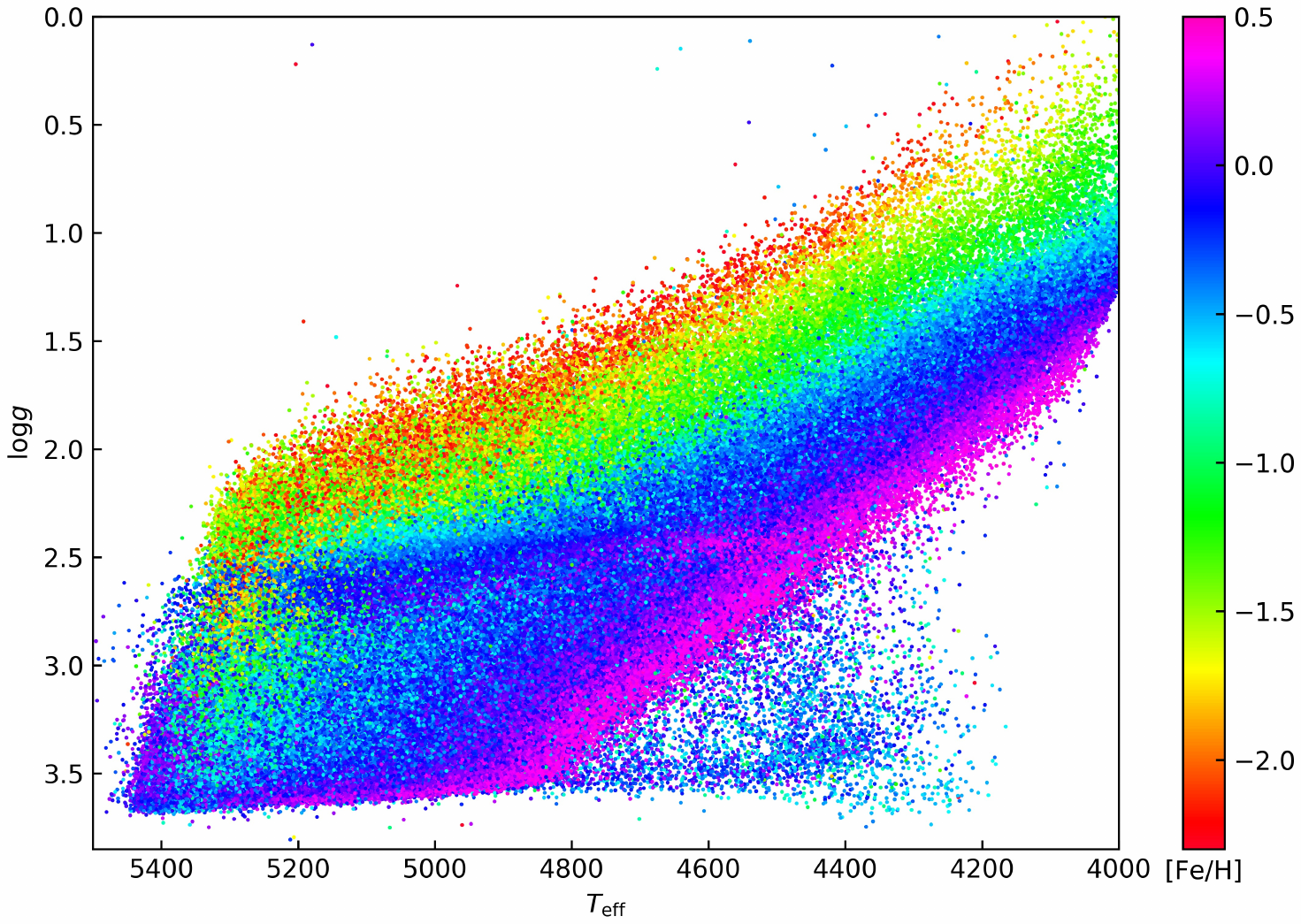}}
\centerline{(c)}
\end{minipage}
\hspace{0pt}
\begin{minipage}[l]{0.46\textwidth}
\leftline{\includegraphics[scale=0.44]{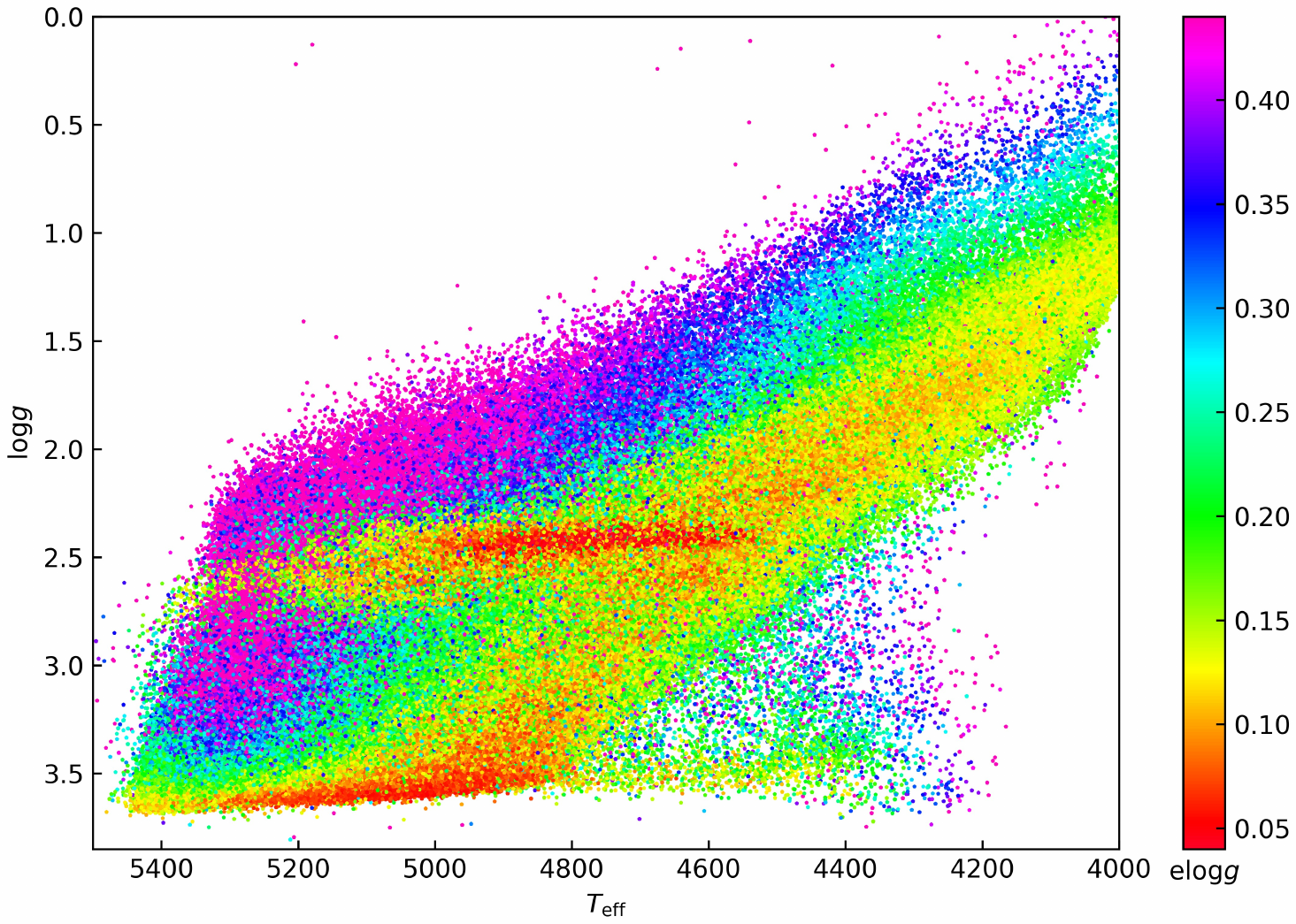}}
\centerline{(d)}
\end{minipage}
\end{center}
\caption{Subplot (a) is HR diagram for APOGEE labels colored by [Fe/H] as a reference. Subplot (b) is HR diagram for neural network predictions of training sample. Subplots (c) and (d) show HR diagrams for predicted log $g$ versus $T_{\textrm{eff}}$ distribution of target sample color-scaled by [Fe/H] and uncertainties of log $g$ respectively.\label{fig:hrd}}
\end{figure}

Figure \ref{fig:hrd} shows distributions of our samples in the stellar parameter space. Subplot (a) presents log $g$ versus $T_{\textrm{eff}}$ from APOGEE color scaled by [Fe/H] as a reference. Subplot (b) presents neural network predicted log $g$ versus $T_{\textrm{eff}}$ of training sample color scaled by predicted metallicity. Subplots (c) and (d) show HR diagrams for predicted log $g$ versus $T_{\textrm{eff}}$ distribution of target sample color-scaled by [Fe/H] and uncertainties of log $g$ respectively. The neural network is highly confident in its predictions along the well-populated parts which is clear from the narrow red clump in subplot (c) and (d). Whenever a label outside the range of the training set is analyzed, the predicted uncertainties are very large too. Thus these large uncertainties can used as a warning flag that they should be treated carefully especially for labels around the border. The distributions of subplots (b), (c) and (d) are very similar to subplot (a) which proves most of our prediction are reliable.

\subsubsection{$\alpha$ elemental abundances}

\begin{figure}[htpb]
\epsscale{.50}
\plotone{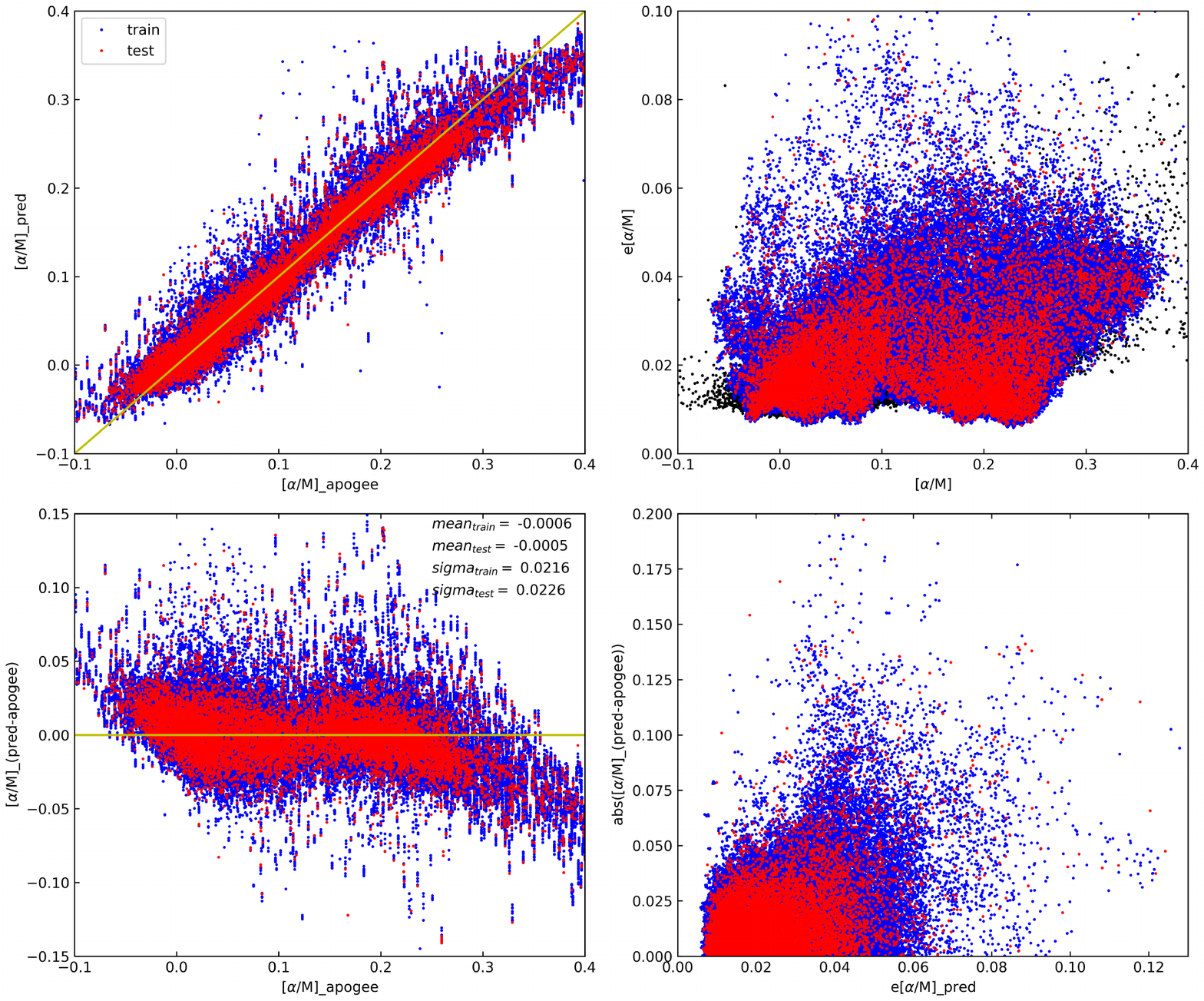}
\caption{Training and test results of [$\alpha$/M] and the meaning of symbols are the same as in figure \ref{fig:f1}.\label{fig:amt}}
%%\end{figure}
%%\begin{figure}[htpb]
%\plotone{afe.png}
%\caption{Training and test results of [$\alpha$/Fe].\label{fig:afet}}
\end{figure}

We not only obtained [$\alpha$/M] as shown in figure \ref{fig:amt}, but also obtained individual $\alpha$ elements such as [Mg/Fe], [Si/Fe], [S/Fe], [Ca/Fe] and [Ti/Fe]. As the bottom-left subplot of figure \ref{fig:amt} shows, the mean values of residuals from both training set and test set are respectively -0.0006 and -0.0005 very close to zero which means the systematic bias is very small. The standard deviation values of both sets are respectively 0.0216 and 0.0226 which means our predicted $\alpha$ abundances of most stars only deviate from APOGEE labels a little. The results of $\alpha$ abundance are better than other individual elements because the overall $\alpha$ abundance have more usable spectral lines than individual elements. Though it looks all good from the left columns of figure \ref{fig:amt}, our suggested range for [$\alpha$/M] is -0.02 to 0.33. Magnesium, silicon, sulfur, calcium and titanium belong to $\alpha$ elements and they have similar features as [$\alpha$/M].

%
%\begin{figure}[htpb]
%\epsscale{.90}
%\plotone{amvsxms.png}
%\caption{Compare our [$\alpha$/M] with \citet{xia16}.\label{fig:amvsxms}}
%\end{figure}
%Moreover, we compared our neural network predicted [$\alpha$/M] with [$\alpha$/M] from \citet{xia16} in figure \ref{fig:amvsxms}. There is a difference between our results that our range of [$\alpha$/M] is larger than theirs. As the yellow diagonal line shows that if a straight line is used to fit the relation of those points then the slope should be larger than one.

\begin{figure}[htpb]
\epsscale{.50}
\plotone{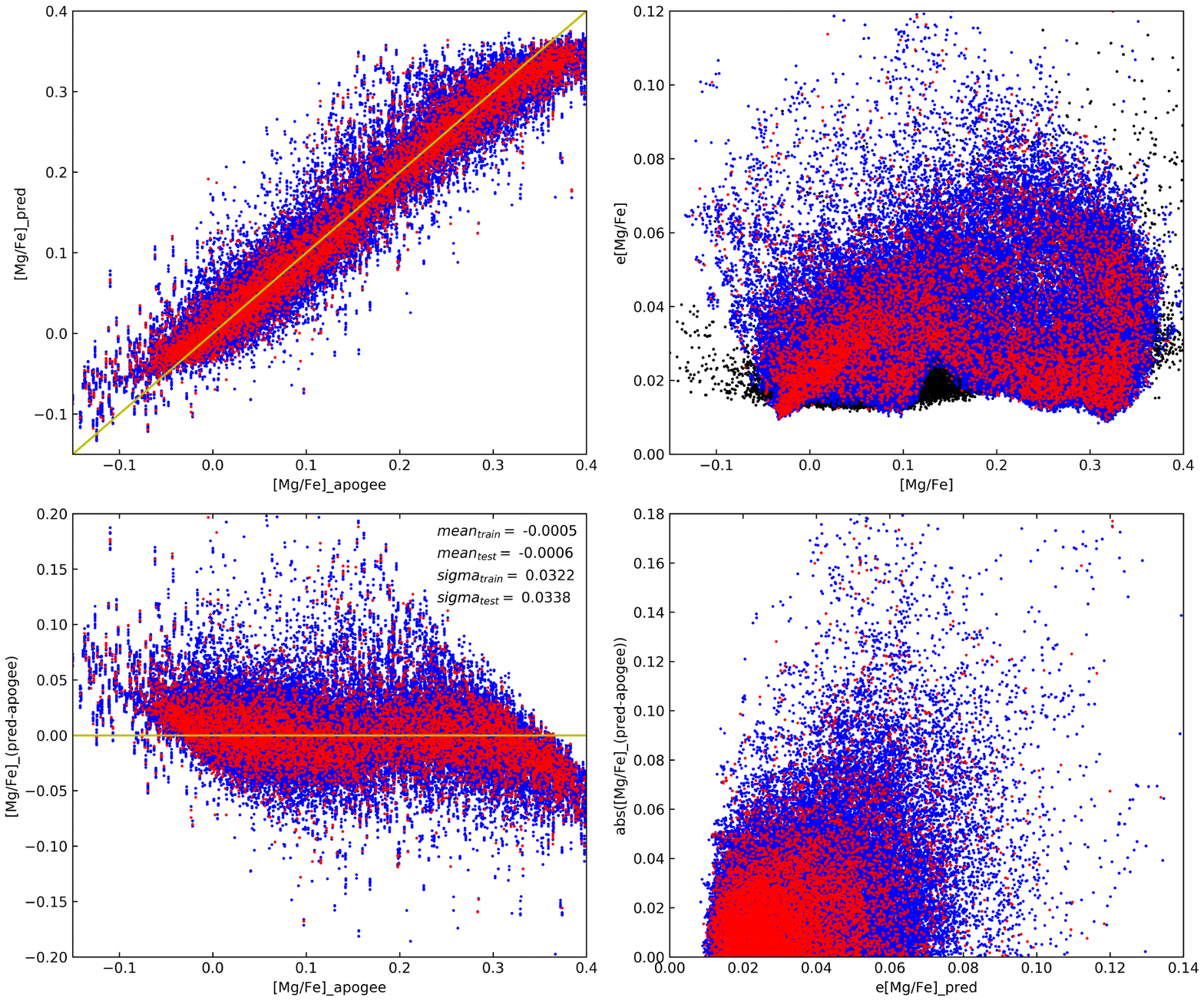}
\caption{Training and test results of [Mg/Fe] and the meaning of symbols are the same as in figure \ref{fig:f1}. \label{fig:mg}}
\end{figure}

Figure \ref{fig:mg} shows our results of [Mg/Fe] for training set and test set. Magnesium have several strong lines even though mixed with other spectral lines and the prediction of magnesium is our best result of all elements. In a low resolution spectra, lines from different elements usually blend together and it is difficult to separate them manually. The advantage of neural network is that it does not need to separate blended lines. If any spectral line plays a role in determining an elemental abundance then it would be adopted by the neural network and the relation does not have to be explicit. From the top left subplot it can be seen that the predicted [Mg/Fe] distributes a relative smaller range than original APOGEE [Mg/Fe] range because of the boundary effect. The original APOGEE [Mg/Fe] of common stellar sample has two density peaks around 0.04 and 0.28 respectively and original APOGEE uncertainties tend to be smaller around peaks than other ranges. The top right subplot and bottom left subplot show our predictions are affected by density distribution of original APOGEE [Mg/Fe]. Though we have already adopted uniformly distributed training sample, this double peaks effect can not be entirely eliminated. Two mean values of residuals of training set and test set are respectively -0.0005 and -0.0006 while two standard deviations of them are respectively 0.0322 and 0.0338. Our recommended highly reliable range for [Mg/Fe] is (-0.05,0.36) judging by top two subplots. The bottom right subplot presents absolute values of residuals distribute with neural network predicted uncertainties. Though there is no clear proportional relation between them in the last subplot, we have to use uncertainties to distinguish good predictions as long as they are positively related.

\begin{figure}[htpb]
\epsscale{.50}
\plotone{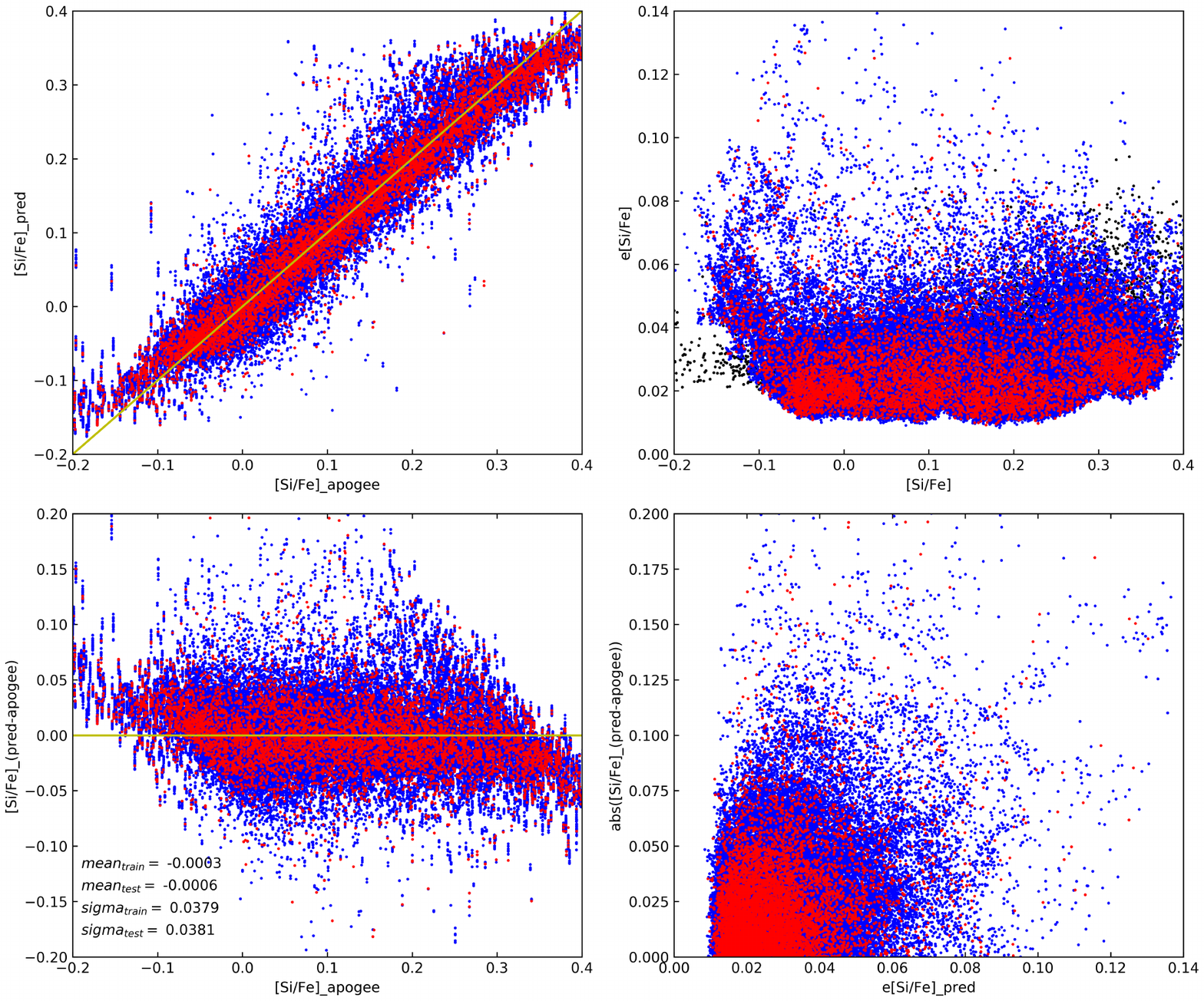}
\caption{Training and test results of [Si/Fe] and the meaning of symbols are the same as in figure \ref{fig:f1}.\label{fig:si}}
\end{figure}

Figure \ref{fig:si} shows our results of [Si/Fe] for training set and test set. Silicon has more spectral line than magnesium but they mixed with other spectral lines more severely. Two subplots at the left column demonstrate our neural network performs well except for the boundary effect that some stars should have been around the boundaries are wrongly predicted tending to the central denser regions. Correspondingly, the top right subplot shows that predicted uncertainties tend to become larger from the center region to boundary regions. Two mean values of residuals of training set and test set are respectively -0.0003 and -0.0006 while two standard deviations of them are respectively 0.0379 and 0.0381. Our suggested highly reliable range for [Si/Fe] is (-0.1, 0.37) and stars outside of the range can still be used as an indicator that they are quite possible to have abundances close to given values. The bottom right subplot still shows a weak positive relation between uncertainties and absolute values of residuals.

\begin{figure}[htpb]
\epsscale{.50}
\plotone{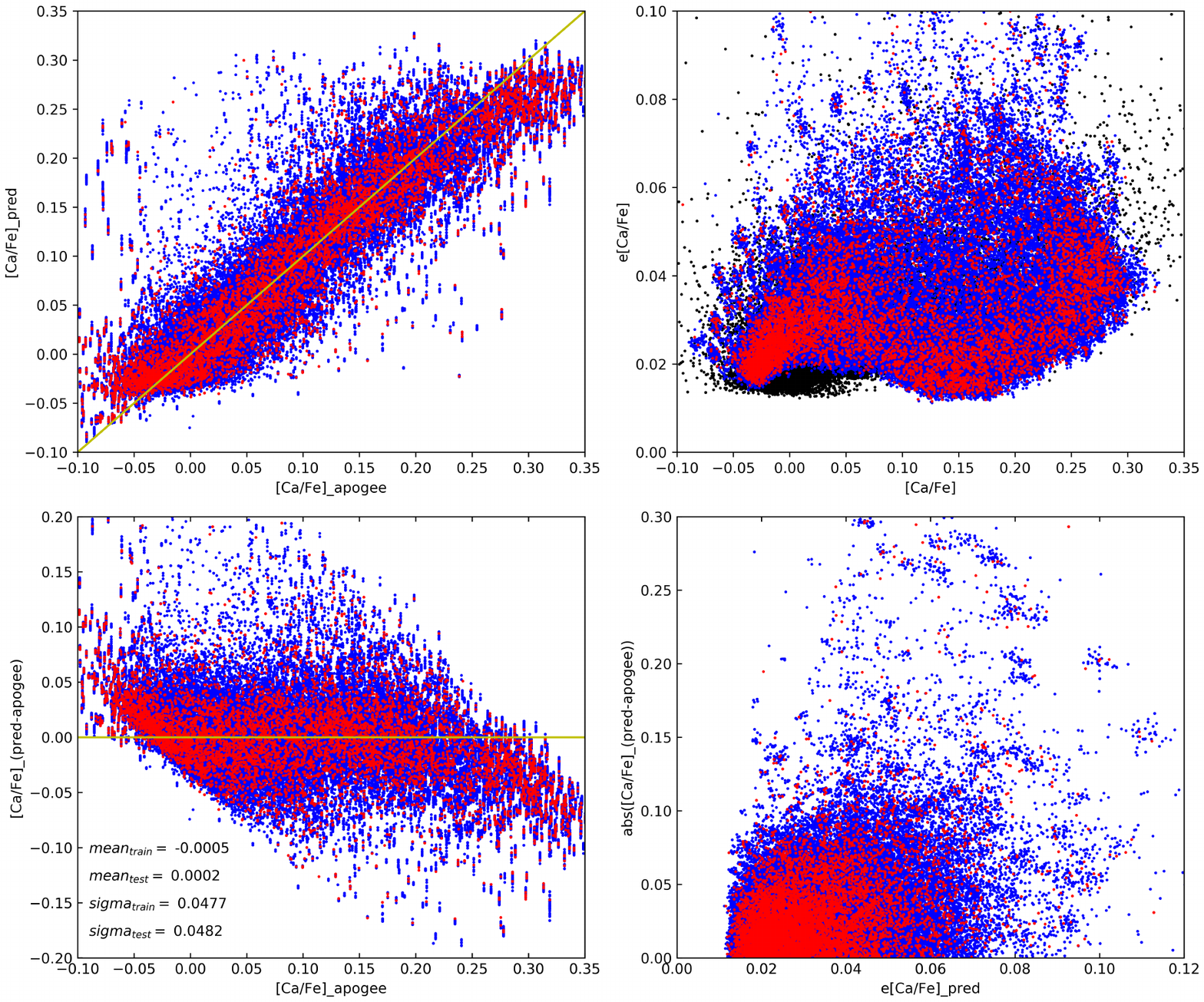}
\caption{Training and test results of [Ca/Fe] and the meaning of symbols are the same as in figure \ref{fig:f1}.\label{fig:ca}}
\end{figure}

Figure \ref{fig:ca} shows our results of [Ca/Fe] for training set and test set. Two top subplots show that the predicted [Ca/Fe] distributes a smaller range than original APOGEE label and our recommended highly reliable range for [Ca/Fe] is (-0.03,0.28). The residual subplot shows obvious boundary effect that stellar calcium abundances at the large end tend to be unestimated and at the small end tend to be overestimated. We think the reason is that not only original star numbers at the boundary take up a small percentage of the common stars sample but also most stellar APOGEE uncertainties are large. One more factor related to the bad result of residuals is that there are only a few Ca\MakeUppercase{\romannumeral1} lines in our spectral wavelength range and they are mixed with other lines. In the lower range of [Ca/Fe], the bad performance might be caused by that spectral lines of calcium are too week for this level of abundances. Two mean values of residuals of [Ca/Fe] from both sets are respectively -0.0005 and 0.0002 while two standard deviations of them are respectively 0.0477 and 0.0482. The top right subplot shows uncertainties of [Ca/Fe] versus [Ca/Fe]. The higher part of [Ca/Fe] range in this subplot shows uncertainties increase calcium abundances and turns back at about 0.3 which leaves a flat edge in the left two subplots. It is obvious both original APOGEE uncertainties and our predicted uncertainties are not horizontally uniform distributed. Thus it is better to treat them locally rather than compare them with stars from the whole range. The bottom right subplot shows absolute residual values are positively related to neural network predicted uncertainties in general.

\begin{figure}[htpb]
\epsscale{.50}
\plotone{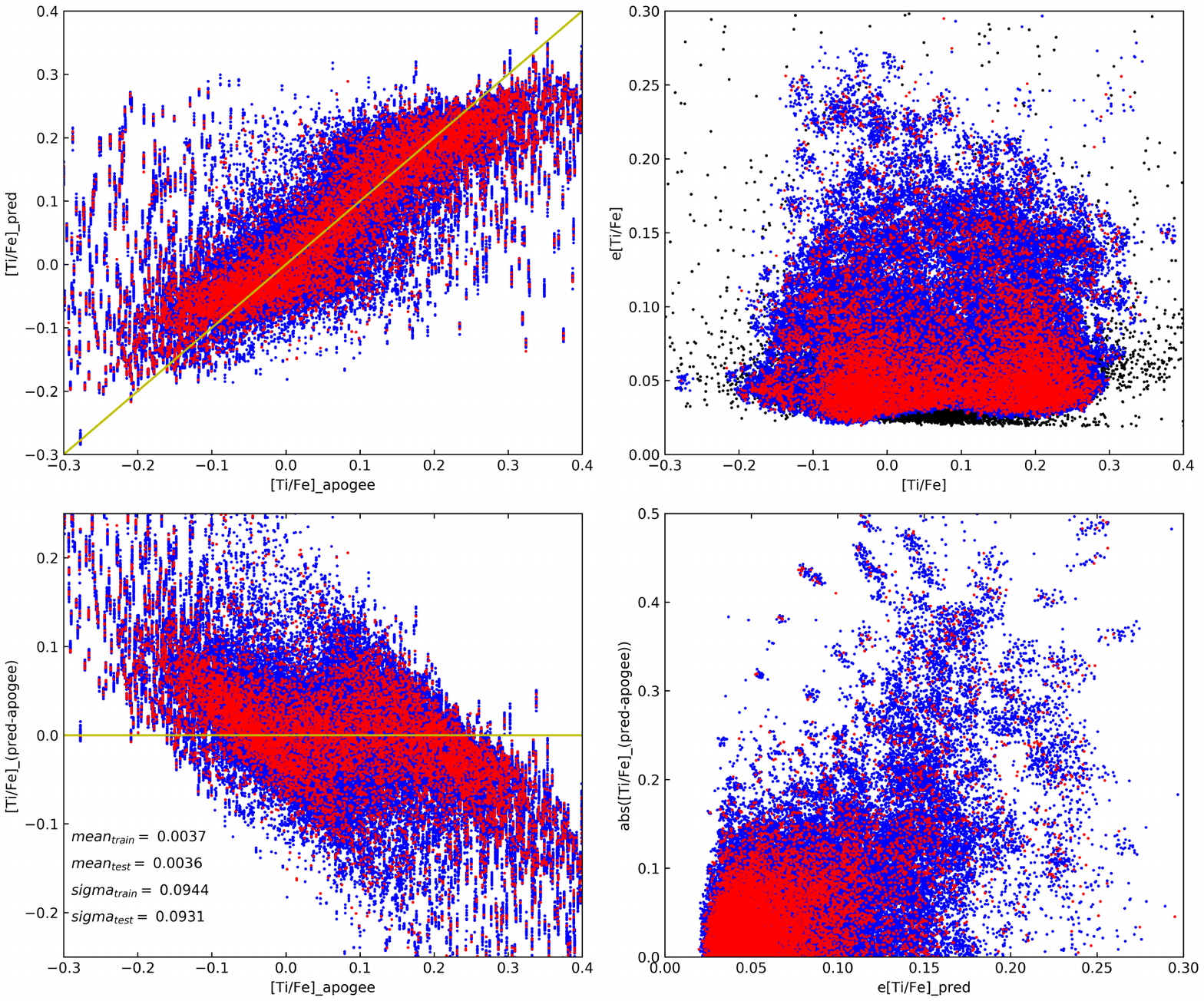}
\caption{Training and test results of [Ti/Fe] and the meaning of symbols are the same as in figure \ref{fig:f1}.\label{fig:ti}}
\end{figure}

Figure \ref{fig:ti} shows our results of [Ti/Fe] for training set and test set. Titanium has more -9999 APOGEE labels of common stellar sample than former three $\alpha$ elements and larger uncertainties than them. Actually APOGEE catalogue provides both [Ti/Fe] and [Ti\MakeUppercase{\romannumeral2}/Fe] and they do not equal to each other. We chose to adopt [Ti/Fe] to represent titanium abundance because it has less -9999 and relatively smaller uncertainties. From top two subplots, it can be seen that the predicted [Ti/Fe] distributes a narrower range than APOGEE label. The residual subplot shows double peaks and boundary effect which not be eliminated by sampling method. Two mean values of residuals of [Ti/Fe] from both sets are respectively -0.0037 and 0.0036. Two standard deviations of training set and test set are respectively 0.0944 and 0.0931 and our subjectively recommended range for [Ti/Fe] is [-0.14,0.23]. Both residuals and predicted uncertainties are large considering the small value range of [Ti/Fe]. Titanium has more spectral lines than other $\alpha$ elements but they all mixed with other elemental lines. Even so, we think our predicted titanium for most stars are good since their original APOGEE uncertainties are large. The last subplot shows most stellar residuals are positively related to predicted uncertainties, thus we can use them to select relatively good prediction when use our data.

\begin{figure}[htpb]
\epsscale{.50}
\plotone{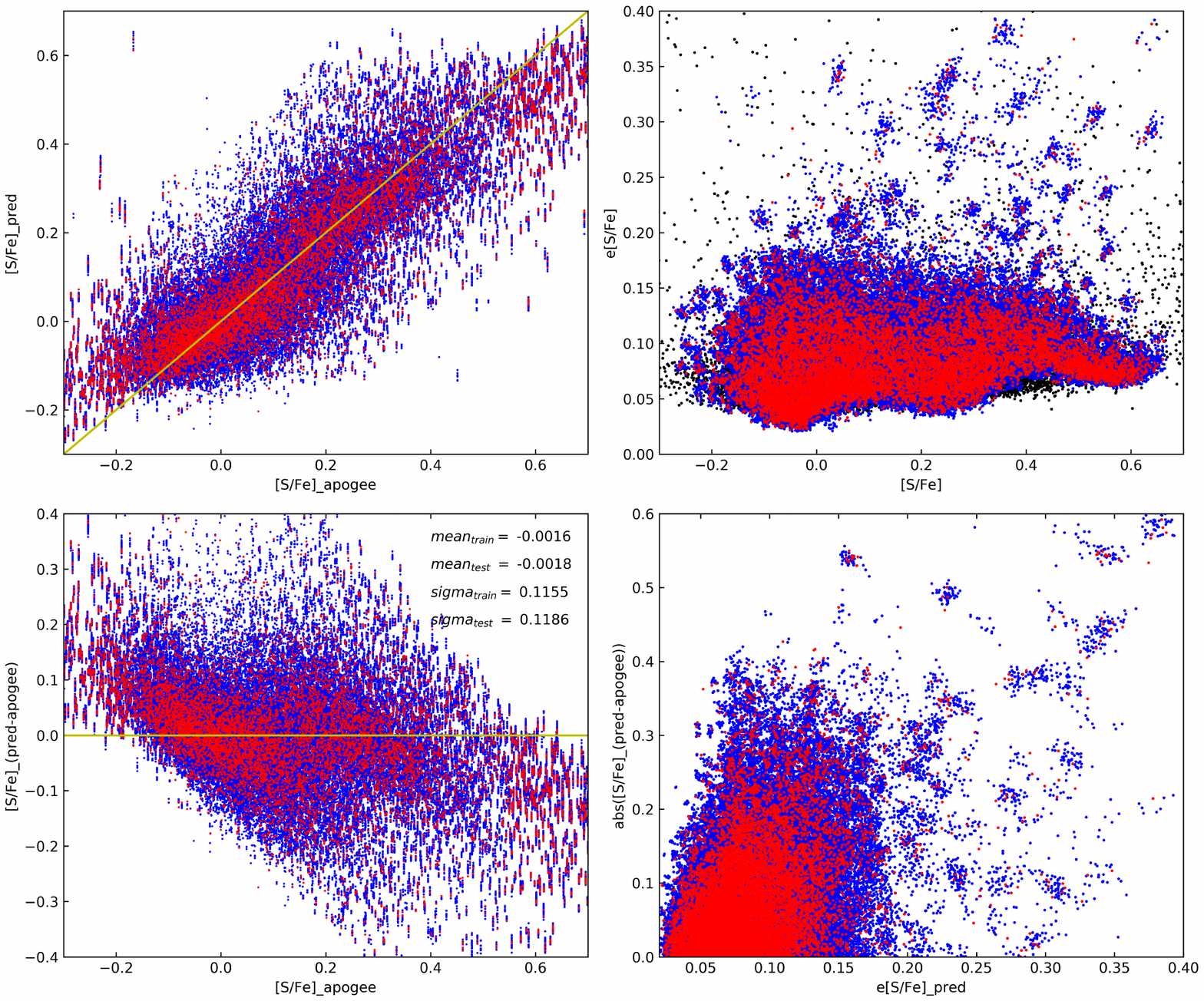}
\caption{Training and test results of [S/Fe] and the meaning of symbols are the same as in figure \ref{fig:f1}.\label{fig:s}}
\end{figure}

Figure \ref{fig:s} shows our results of [S/Fe] for training set and test set. Sulfur has more -9999 than former four $\alpha$ elements and harder to obtain abundances even for high resolution spectra. Like former elements, the top left subplot shows cross validation with APOGEE labels and it also has double peak and boundary problem. We think the flat bottom boundary is related to sulfur line are too weak there and the flat top boundary might be related metallicity. The bottom left subplot presents the residual distribution of [S/Fe] and our subjectively recommended range for [S/Fe] is (-0.15, 0.5). Two mean values of residuals of [S/Fe] from both training set and test set are respectively -0.0016 and -0.0018 while two standard deviations of them are respectively 0.1155 and 0.1186. The top right subplot presents uncertainties of [S/Fe] versus [S/Fe] and those large uncertainty clumps either come from border region or related metal poor stars. The bottom right subplot presents absolute values of residuals versus predicted uncertainties of [S/Fe] and they are positively related.

\subsubsection{Carbon, nitrogen and oxygen}

\begin{figure}[htpb]
\epsscale{.50}
\plotone{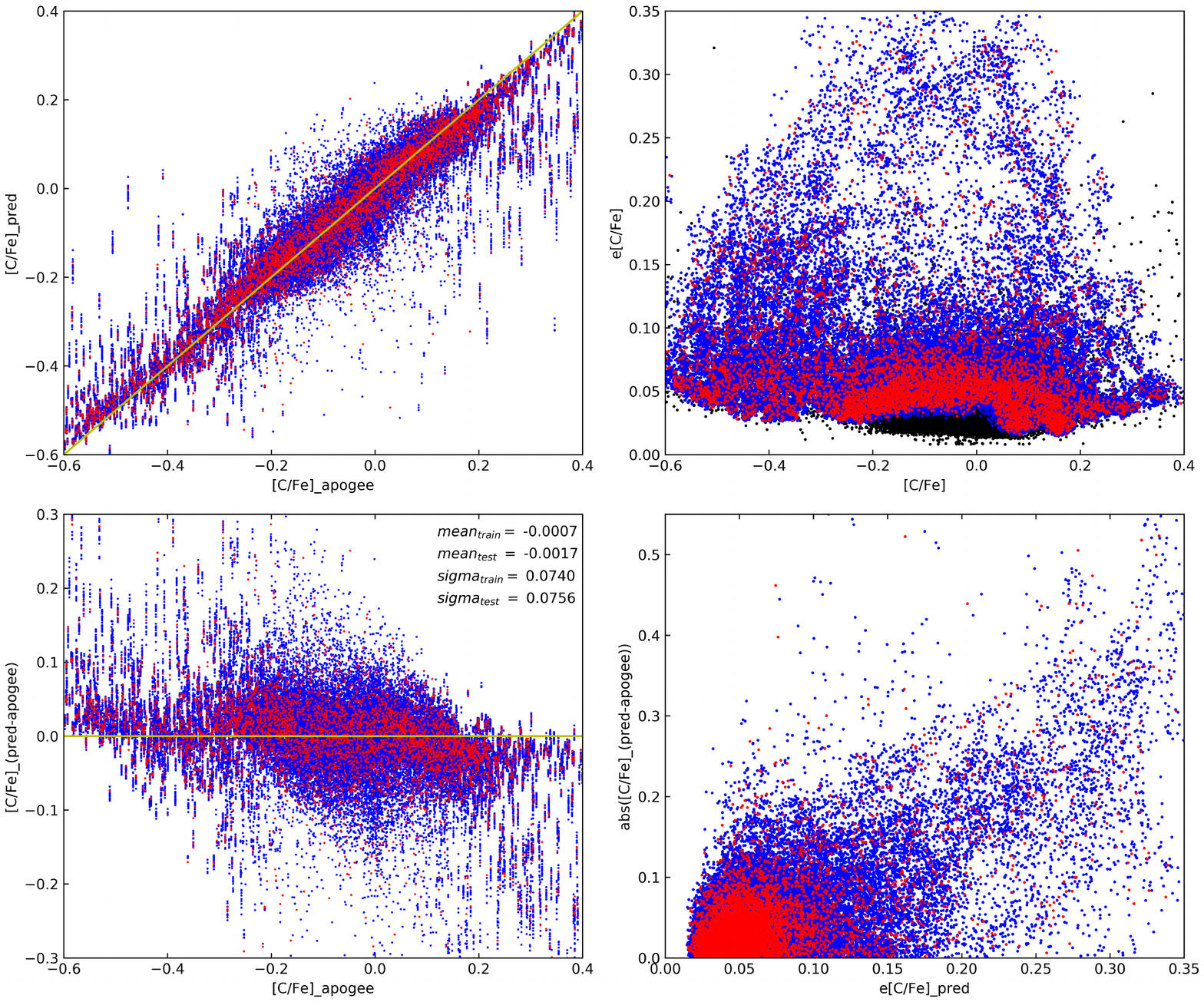}
\caption{Training and test results of [C/Fe] and the meaning of symbols are the same as in figure \ref{fig:f1}.\label{fig:c}}
\end{figure}

Figure \ref{fig:c} shows our results of [C/Fe] for training set and test set. Obtaining accurate carbon abundance is not easy even for high resolution spectra. In our spectra carbon only have some weak spectral lines at G-band which usually mixed with nitrogen lines. Moreover our spectra have relatively low SNR at the short wavelength region. Our common stars sample is mainly consist of giants and carbon along with nitrogen and oxygen of many stars have already evolved which makes the situation worse. Nevertheless, carbon and nitrogen are important elements which contain age information. As former labels, the top left subplot shows cross-validation with reference APOGEE [C/Fe] and the subplot under it shows residuals distribute with APOGEE [C/Fe]. Two mean values of residuals of [C/Fe] from both training set and test set are respectively -0.0007 and -0.0017 while two standard deviations of them are respectively 0.074 and 0.0756. According to all subplots in figure \ref{fig:c}, our recommended range for [C/Fe] is (-0.5,0.25) when use our data. For most stars, results are reasonable that there is almost no systematic bias and most residuals are small. Those stars with not very good predictions might be due to weak and mixed spectral lines and low SNR especially for metal poor stars. The top right subplot shows the uncertainties of [C/Fe] distribute with [C/Fe]. Those stars with large uncertainties shows obvious tendency to incline towards central region and uncertainties can be used to distinguish them from stars supposed to be there. The bottom right subplot shows absolute values of residuals versus predicted uncertainties and uncertainties will be used to cut off bad predictions.

\begin{figure}[htpb]
\epsscale{.50}
\plotone{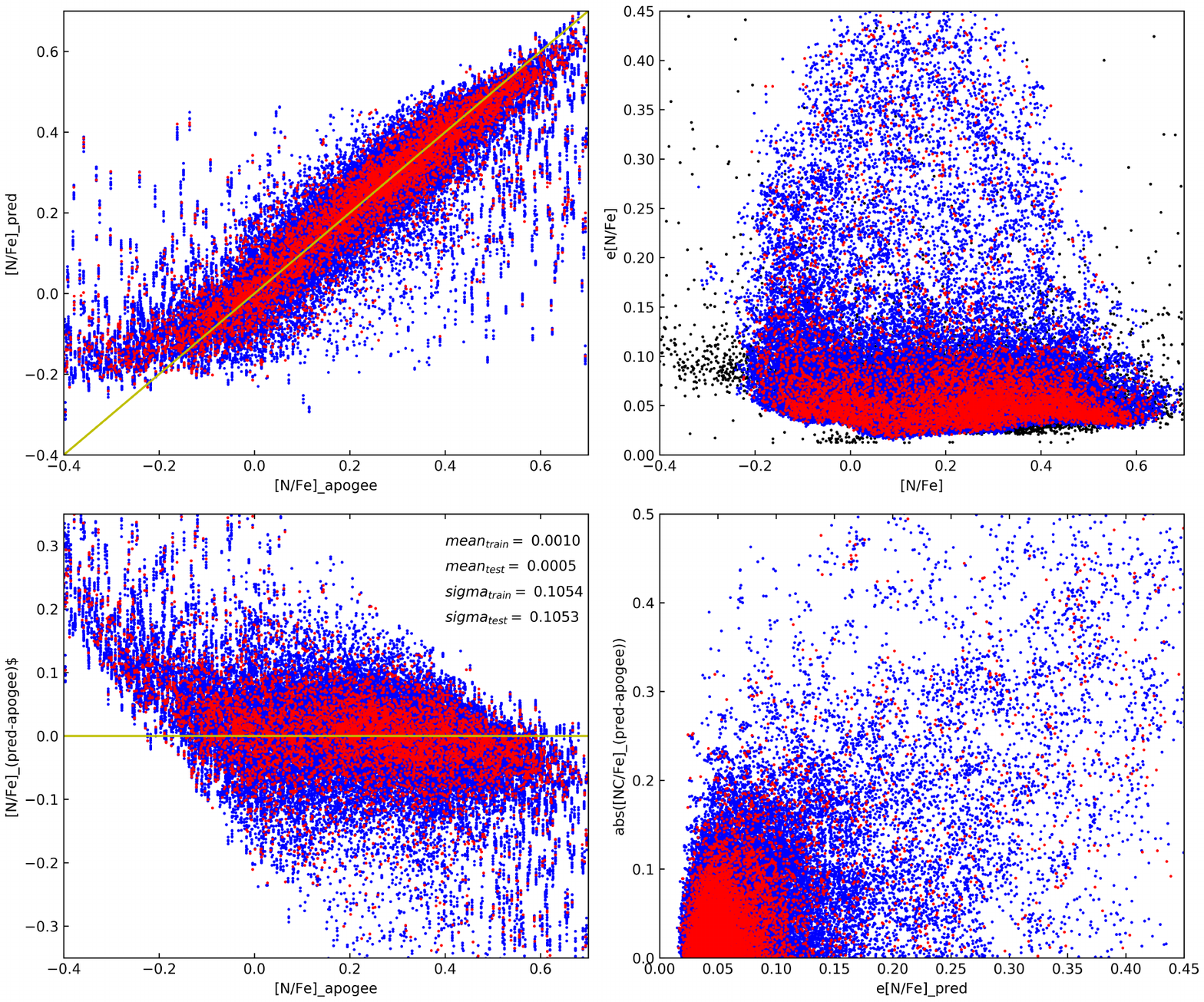}
\caption{Training and test results of [N/Fe] and the meaning of symbols are the same as in figure \ref{fig:f1}.\label{fig:n}}
\end{figure}

Figure \ref{fig:n} shows our results of [N/Fe] for training set and test set. In the first subplot, the bottom bound of predicted [N/Fe] seams flat and may be because spectral lines are too weak at this level of abundance to be treated as effective feature by the network. The top right subplot presents the distribution of uncertainties of [N/Fe] versus [N/Fe]. For stars with large uncertainties they show similar tendency as in the corresponding subplot of carbon figure \ref{fig:c}. Our recommended range for [N/Fe] is (-0.2, 0.6) and residuals of most stars are small compare to such a wide range of [N/Fe]. Two mean values of residuals of [N/Fe] from both training set and test set are respectively 0.001 and 0.0005 while two standard deviations of them are respectively 0.1053 and 0.1054. The bottom right subplot shows obvious positive relation between absolute values of residuals and neural network predicted uncertainties.

\begin{figure}[htpb]
\epsscale{.50}
\plotone{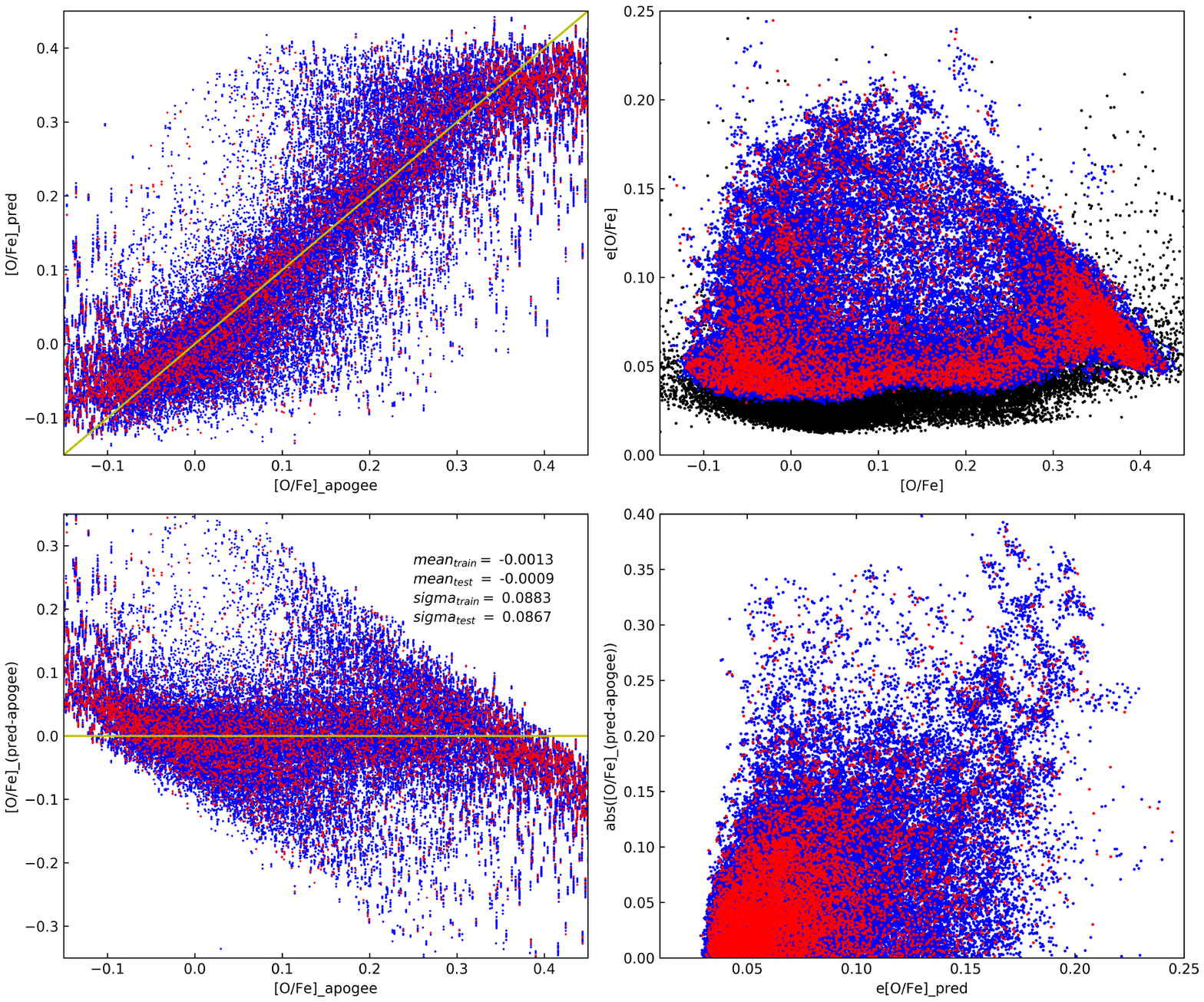}
\caption{Training and test results of [O/Fe] and the meaning of symbols are the same as in figure \ref{fig:f1}.\label{fig:o}}
\end{figure}

Figure \ref{fig:o} shows our results of [O/Fe] for training set and test set. Oxygen abundance is very difficult to obtain even for high resolution spectra and there is no independent or strong oxygen line in our low resolution spectra. The boundary effect is obvious in the first subplot and our recommended range for [O/Fe] is (-0.05, 0.35). The bottom left residual subplot shows two flat boundaries and we think they are not only induced by boundary effect but also related to metal poor stars. The top right subplot presents the distribution of uncertainties of [O/Fe] versus [O/Fe]. For stars with large uncertainties they show similar tendency as former two elements. Judging by the right end of the top right subplot, the top boundary of oxygen abundance may be over-fitted a little because too many stars clumped at that region. As shown in the bottom left subplot, mean values of residuals of [O/Fe] from training set and test set are respectively -0.0013 and -0.0009 while standard deviations of them are respectively 0.0883 and 0.0867. The last subplot shows obvious positive relation between absolute values of residuals and neural network predicted uncertainties.

\subsubsection{Odd Z elements}

\begin{figure}[htpb]
\epsscale{.50}
\plotone{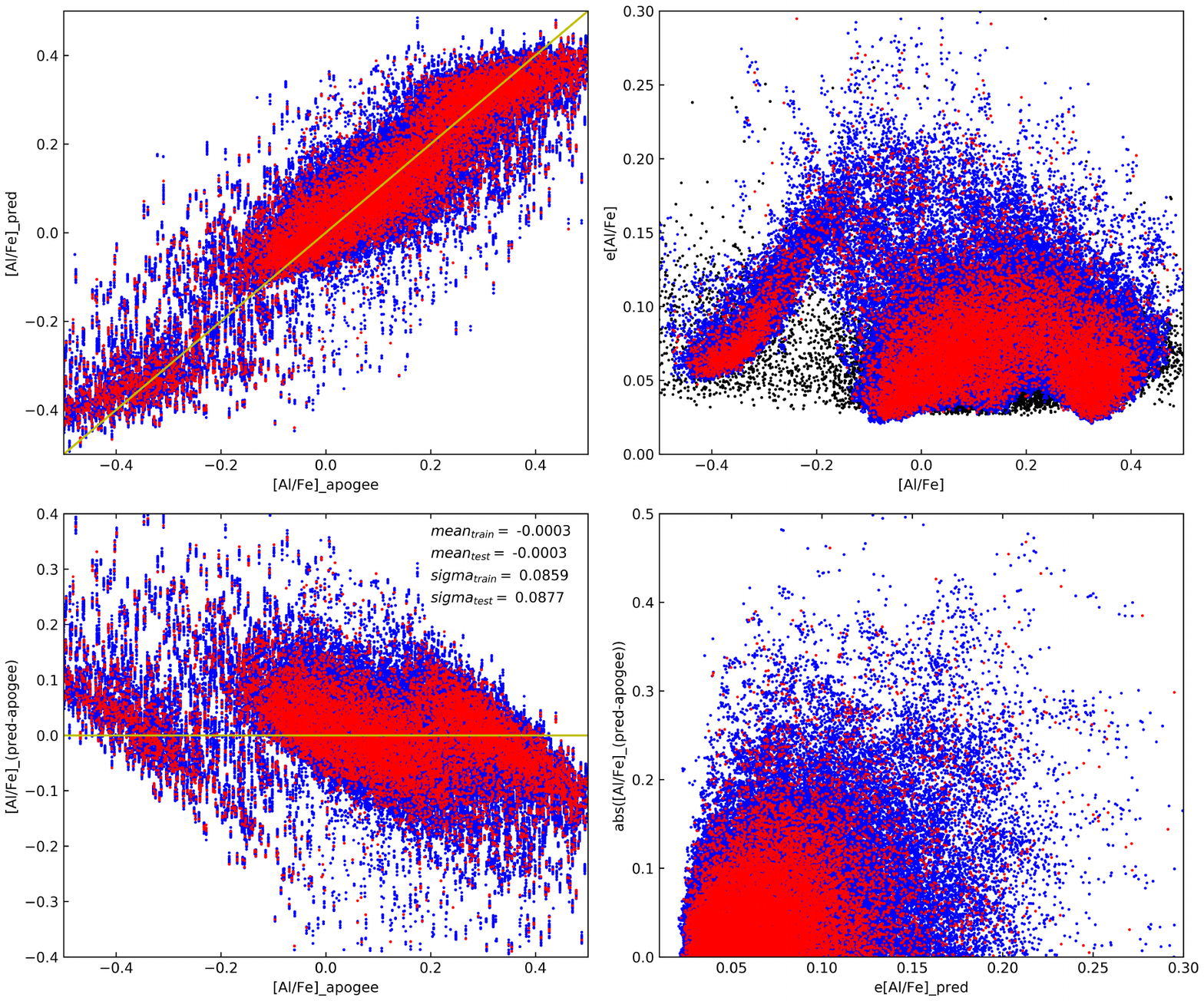}
\caption{Training and test results of [Al/Fe] and the meaning of symbols are the same as in figure \ref{fig:f1}.\label{fig:al}}
\end{figure}

Figure \ref{fig:al} shows our results of [Al/Fe] for training set and test set. Aluminum abundances show three parts distributions because number density of original APOGEE [Al/Fe] in the common stellar sample has three peaks and APOGEE uncertainties clump three groups. From two subplots in the left column, our recommended range for [Al/Fe] is (-0.4, 0.4). Judging by the top right subplot and bottom left subplot, some stars deviate from their right positions towards center region and come with large uncertainties. Mean value of residuals of two sets are both -0.0003 and there is basically no systematic bias between our predictions and APOGEE labels. Standard deviations of training set and test set are respectively -0.0859 and 0.0877 and most stellar residuals are in acceptable range.

\begin{figure}[htpb]
\epsscale{.50}
\plotone{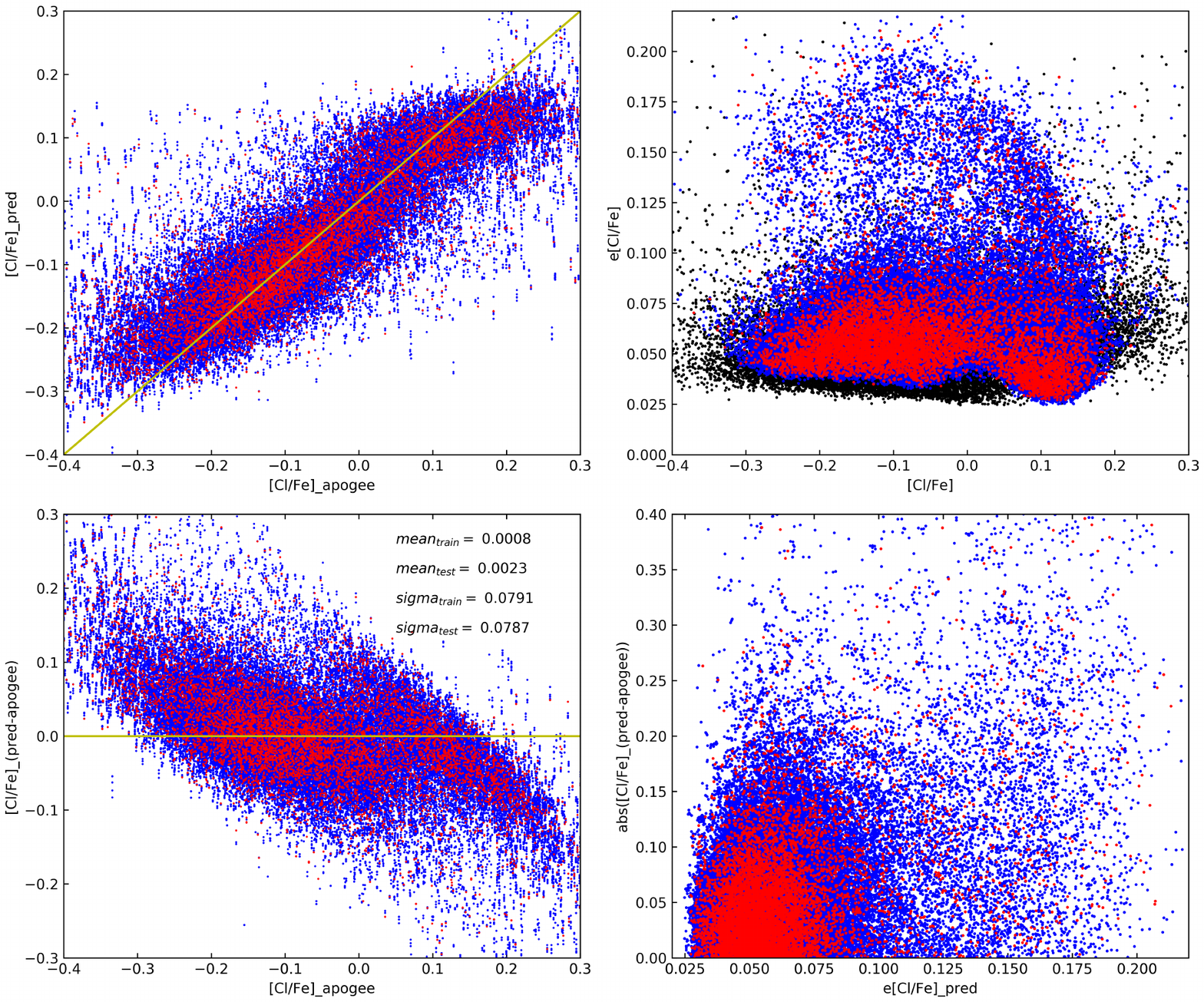}
\caption{Training and test results of [Cl/Fe] and the meaning of symbols are the same as in figure \ref{fig:f1}.\label{fig:cl}}
\end{figure}

Figure \ref{fig:cl} shows our results of [Cl/Fe] for training set and test set. Chlorine is a seldom analysed element because its abundance is very hard to obtain. From the top left subplot, our predicted [Cl/Fe] range is smaller than real range due to boundary effect. Combining top right subplot and bottom left subplot, it can be seen that some stars from top boundary region of [Cl/Fe] are underestimated and they have larger predicted uncertainties than those rightly predicted stars. Those wrongly predicted stars are mostly related to metal poor stars and our recommended range for [Cl/Fe] is (-0.25,0.15). Mean values of residuals of [Cl/Fe] from training set and test set are respectively 0.0008 and 0.0023 while standard deviations of them are respectively 0.0791 and 0.0787. The bottom right subplot shows distribution of absolute values of residuals versus predicted uncertainties.

\subsubsection{Iron group elements}

Manganese, iron and nickel are iron group elements and they are supposed to be mainly produced by type \MakeUppercase{\romannumeral1}a supernovae whose predecessors are small mass stars that lives a relatively long life. On the contrast, short lived massive stars are the major source of $\alpha$ elements and the light odd Z elements. They return synthesised material to interstellar medium through stellar winds and Type \MakeUppercase{\romannumeral2} supernovae explosion which is much faster.

\begin{figure}[htpb]
\epsscale{.50}
\plotone{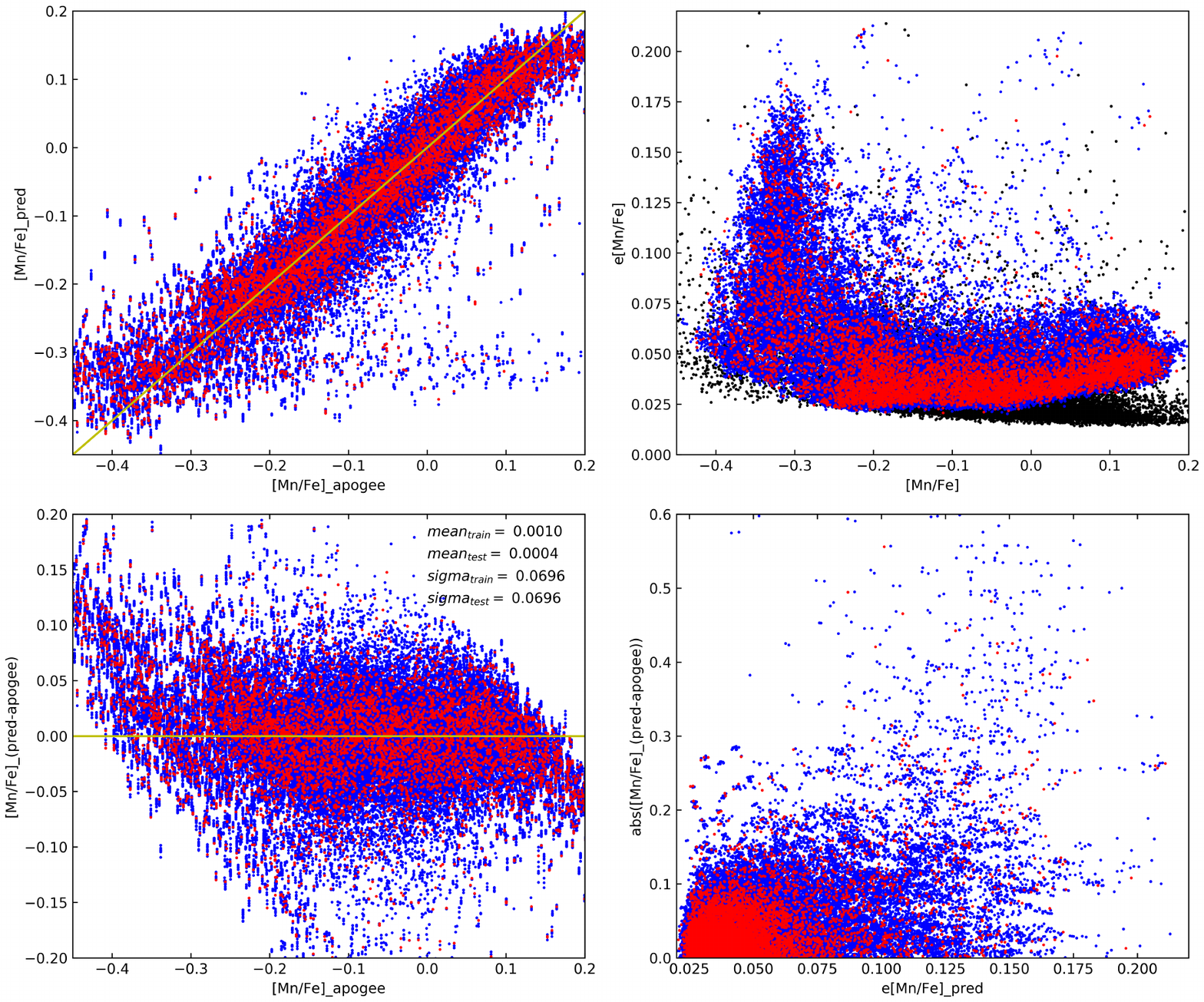}
\caption{Training and test results of [Mn/Fe] and the meaning of symbols are the same as in figure \ref{fig:f1}.\label{fig:mn}}
\end{figure}

Figure \ref{fig:mn} shows our results of [Mn/Fe] for training set and test set. The manganese abundance is also very hard to obtain because it only has week mixed lines. The top left subplot shows some stars are wrongly predicted around -0.3 and our recommended range for [Mn/Fe] is (-0.28,0.16). Many wrong predictions belong to metal poor stars and they have large predicted uncertainties as can be seen in the top right subplot. For metal poor star, those weak mixed manganese lines that are treated by neural network as features are dominated by iron lines. As show in the subplots of manganese in figure \ref{fig:jg}, those blue and pink dots show an approximately linear relation between [Mn/Fe] and [Fe/H] and those stars are wrongly estimated to values around [Mn/Fe]$=$-0.3. Mean values of residuals of [Mn/Fe] from training set and test set are respectively 0.0010 and 0.0004 while standard deviations of them are both 0.0696. So there is basically no systematic bias and most other stars are still usable. Even for stars around -0.3, uncertainties can be used to select good predictions. In the bottom right subplot, positive related relation between absolute values of residuals and predicted uncertainties is clear for large errors.

\begin{figure}[htpb]
\epsscale{.50}
\plotone{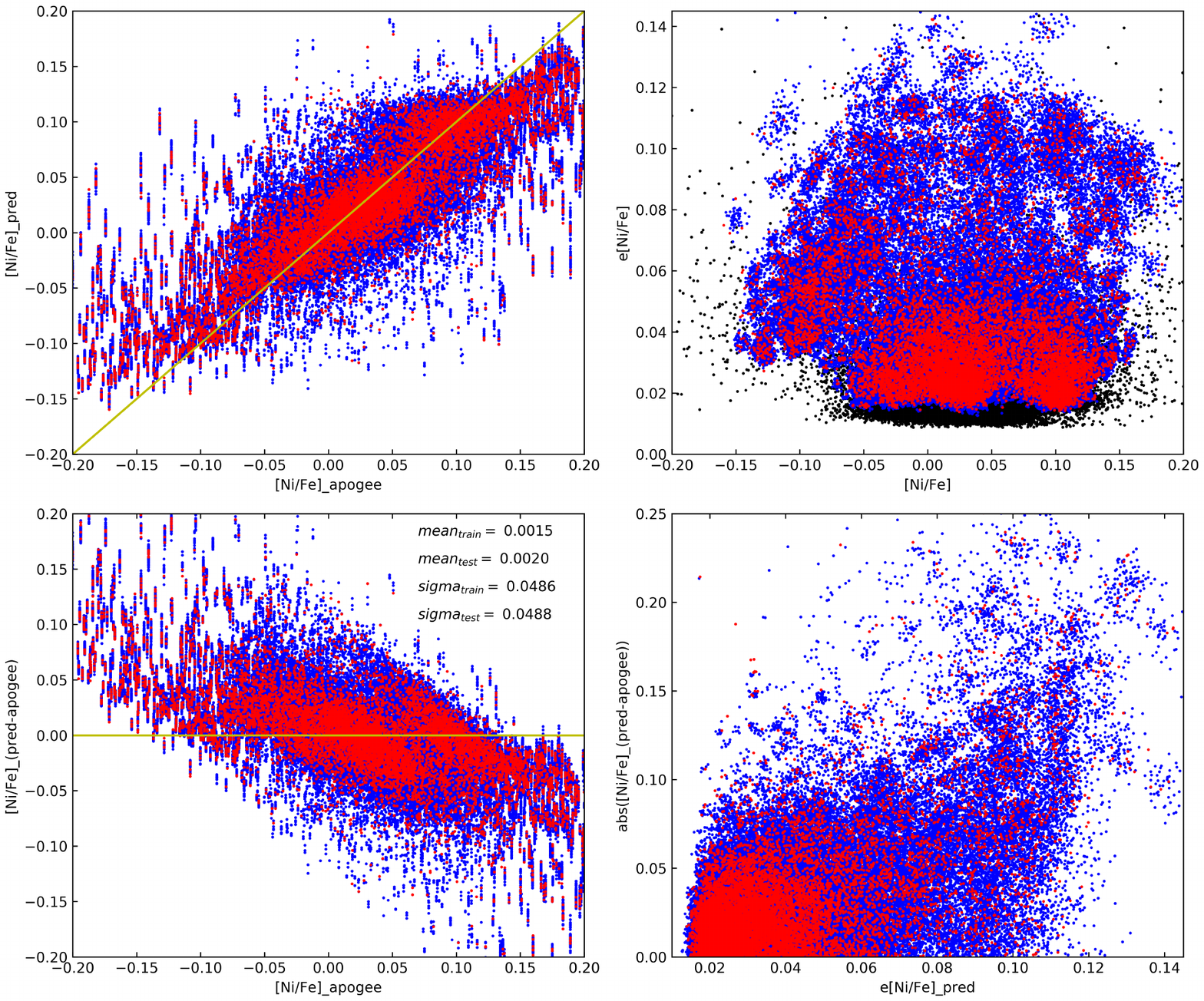}
\caption{Training and test results of [Ni/Fe] and the meaning of symbols are the same as in figure \ref{fig:f1}.\label{fig:ni}}
\end{figure}

Figure \ref{fig:ni} shows our results of [Ni/Fe] for training set and test set. Nickel element has many lines but most of them are mixed with other lines in our low resolution spectra. Both residuals and predicted uncertainties are not very large for most stars. However, value range of [Ni/Fe] is not very large too and our recommended range for our predicted [Ni/Fe] is (-0.07,0.13). As shown in the top right subplot, stars at the boundary have relatively large uncertainties and tend to tilt towards the center region for top part. That is some stars from the boundary ranges are wrongly estimated towards denser ranges and the more wrong the larger uncertainties are. The bottom right subplot shows clear positive relation between absolute values of residuals and neural network predicted uncertainties.

\section{RESULT AND DISCUSSION}

In a word, the neural network is good enough for most stars even though it performs poorly at boundaries. Neural network works differently from high order fitting and more similar to classification. We suppose it treats every star label as a class and find similarities and differences to obtain weights. It uses weights to guess rather than fitted function to interpolate. This method is better at dense regions in label range where discrete values can approximately substitute continuous distribution. However for those very sparse regions in the label range, classification can not get decent values. Such as it is, the predicted values in the sparse regions can still be used as an indicator qualitatively.

\begin{figure}[htpb]
\epsscale{.80}
\plotone{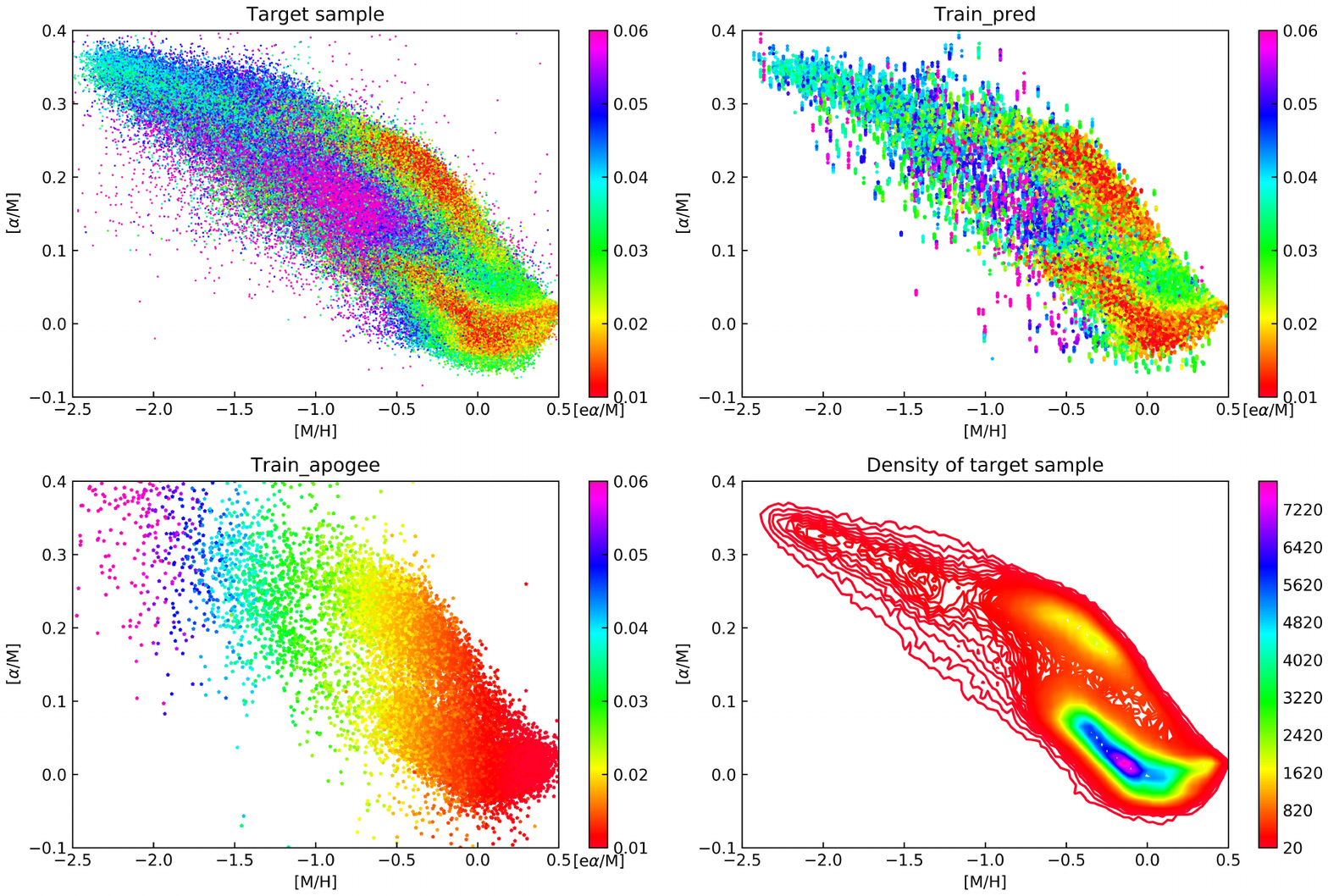}
\caption{$\alpha$ abundances versus metallicity color scaled by uncertainties of [$\alpha$/M] for target sample and training set and a density plot for target sample.\label{fig:ammh}}
\end{figure}

Figure \ref{fig:ammh} shows $\alpha$ abundances versus metallicity color scaled by uncertainties of [$\alpha$/M] and a density plot for target sample. In the top right subplot, four parts with small uncertainties are obvious and they are thin disk, thick disk, inner halo and out halo. Those violet points clustered at where stars are sparse in reality and blue points are distributed around boundaries. This subplot proves most our prediction are useful when used with uncertainties for different stellar components. The top right subplot and bottom left subplot show respectively predicted labels and APOGEE labels for training set as a comparison. The distribution of target sample is very similar to the distribution of predictions of training set. The APOGEE uncertainties of [$\alpha$/M] is positive related to [M/H] while predicted uncertainties of [$\alpha$/M] also affected by density distribution of APOGEE labels. The last contour subplot shows the density distribution of target sample and it shows most stars of our target sample are disk stars. There is similarity between distributions of uncertainties and density on the whole that most dense regions have small predicted uncertainties.

\begin{figure}[htpb]
\plottwo{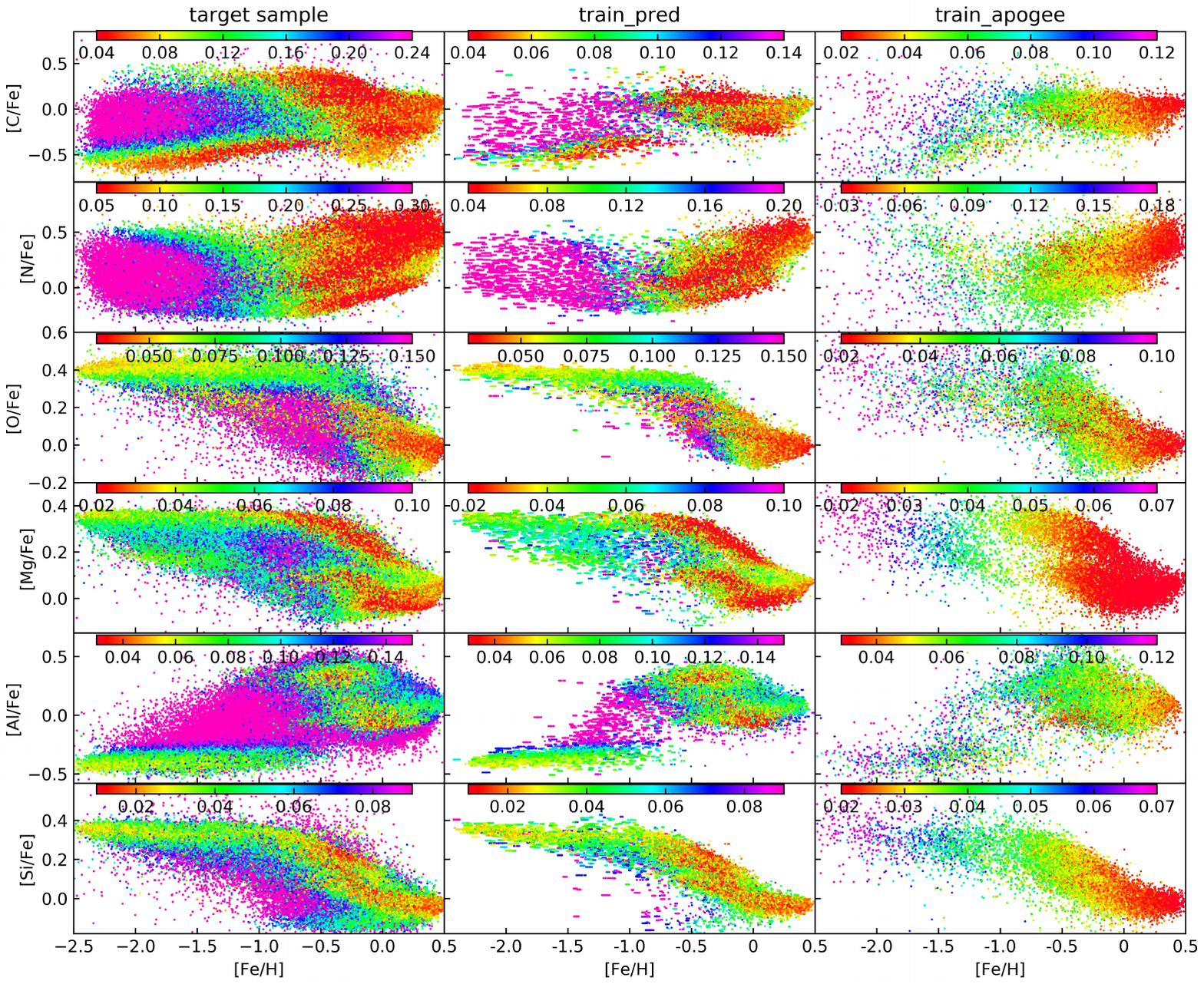}{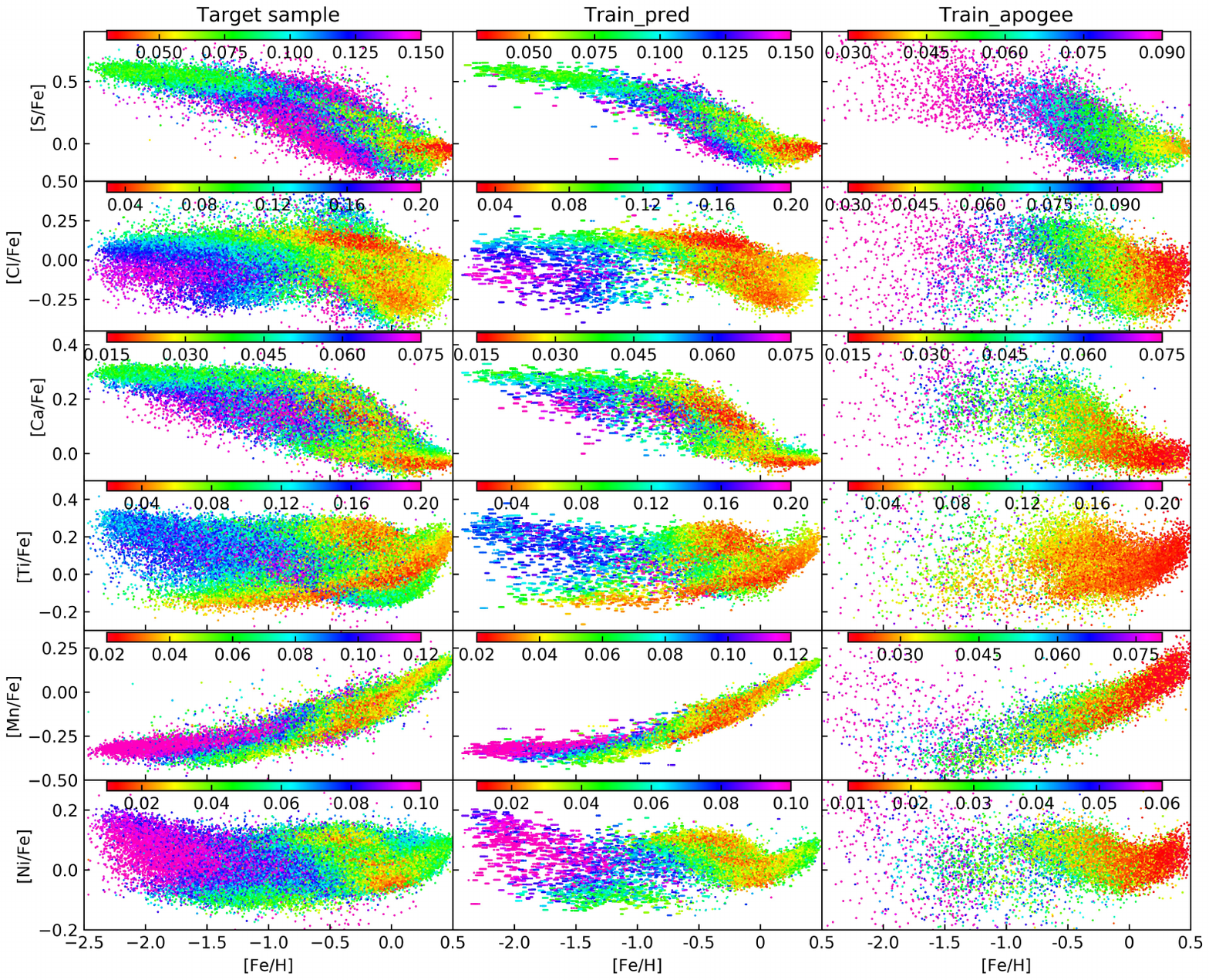}
\caption{Results of twelve elemental abundances versus metallicity color scaled by uncertainties.\label{fig:jg}}
\end{figure}

Figure \ref{fig:jg} shows results of twelve elemental abundances versus metallicity color scaled by uncertainties. The left column and center column respectively present results of neural network predictions of target sample and training sample while the right column presents APOGEE labels of training sample as a comparison. Color bars in the top of each subplot represent uncertainties of each elemental abundances. On the whole, the distributions of our predicted labels are similar to that of APOGEE labels and there are a little differences between them for different elements. The original APOGEE uncertainties of all elements monotonically increase as metallicity becomes poorer while our predicted uncertainties are not only determined by original APOGEE uncertainties but also by density distributions of labels. Stars with small APOGEE uncertainties are mainly more trusted by neural networks and predicted uncertainties are small too. Stars with relatively small uncertainties such as red, orange and green points as a whole take up similar regions between our predictions and APOGEE labels. However those regions occupied by blue and purple points in APOGEE labels subplot have different prediction distributions and mainly in the metal poor region. Comparing with APOGEE labels distribution, some elemental abundances have smaller uncertainties at boundary than they should be which may be caused by overfitting in the metal poor region. For oxygen, sulfur and manganese, the approximately linear relation between elemental abundance and metallicity for metal poor stars may be due to their lines are dominated by mixed iron lines and too weak to provide useful information. Such as it is, we can choose to use relatively reliable predictions to do researches.

\subsection{Elemental abundances distributions in the (R, $V_{\phi}$) plane}

we cross-matched our LAMOST target sample with Gaia DR2 \citep{gai18} catalog and got 1 058 850 common stars. Stars with $\sigma_{\varpi}/\varpi > 0.2$ or $\varpi <= 0$ have been removed. After took out duplicated sources (mainly from LAMOST), 657 561 stars left. The cross-matched sample is mainly composed of giant stars because the common stellar sample is mainly composed of giant stars. With proper motions and parallax from Gaia DR2 and radial velocity along sight from LAMOST DR5, kinematical parameters are calculated in the cylindrical coordinate with origin at the galactic center and \textit{z} axis points to the North Galactic Pole. $V_{\phi}$ is the azimuthal velocity a star rotates around the galactic center while \textit{R} is its galactic radius and $V_R$ is the velocity component along \textit{R}. The distance from the Sun is directly calculated as 1/parallax. The peculiar velocities of the Sun with respect to the Local Standard of Rest are respectively (11.1, 12.24, 7.25) km s$^{-1}$ \citep{sch10}. The sun was placed at $z=0.014$ kpc \citep{bin97} and $R=8.34$ kpc with the circular rotation speed $ V_{c}=240$ km s$^{-1}$ \citep{rei14}.

\begin{figure}[htpb]
\epsscale{.8}
\plotone{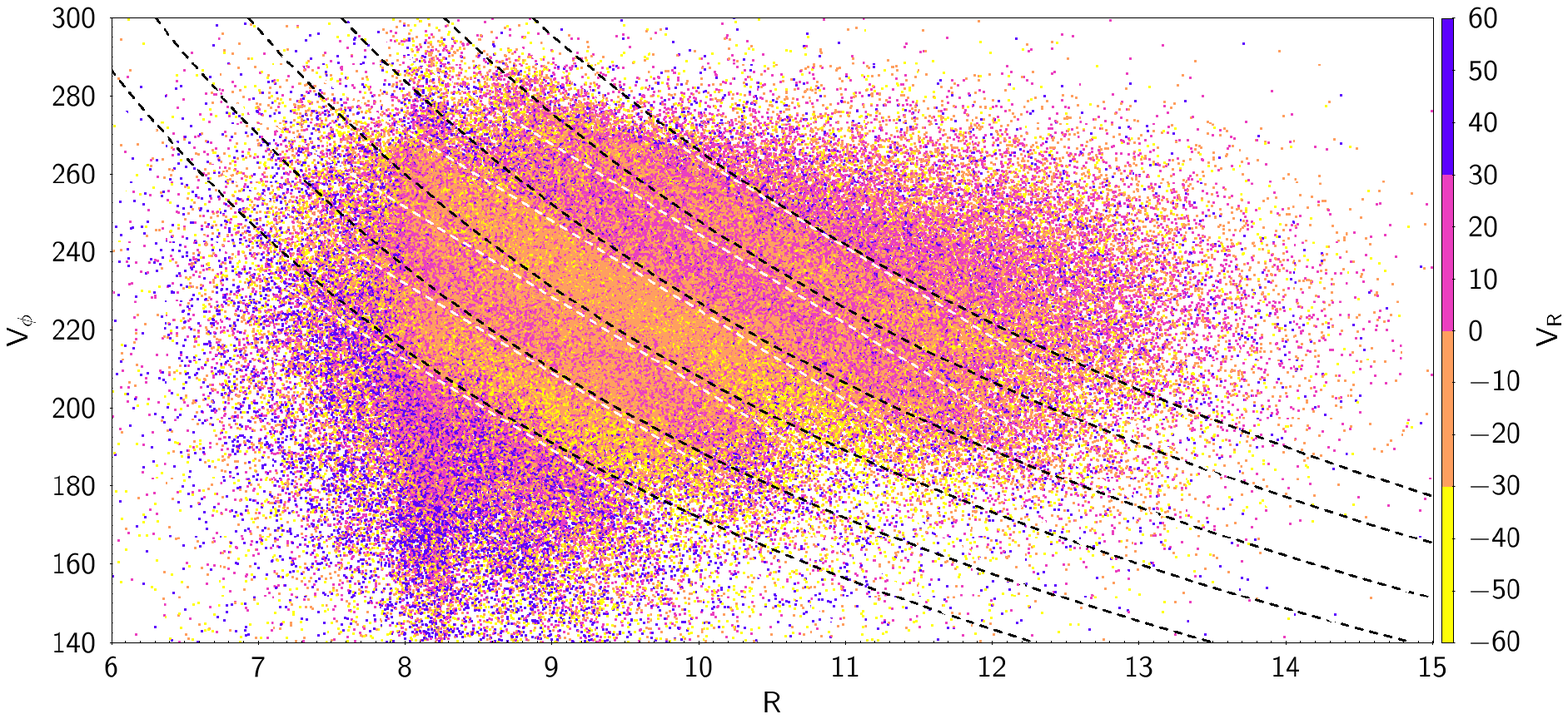}
\hspace{0pt}
\plotone{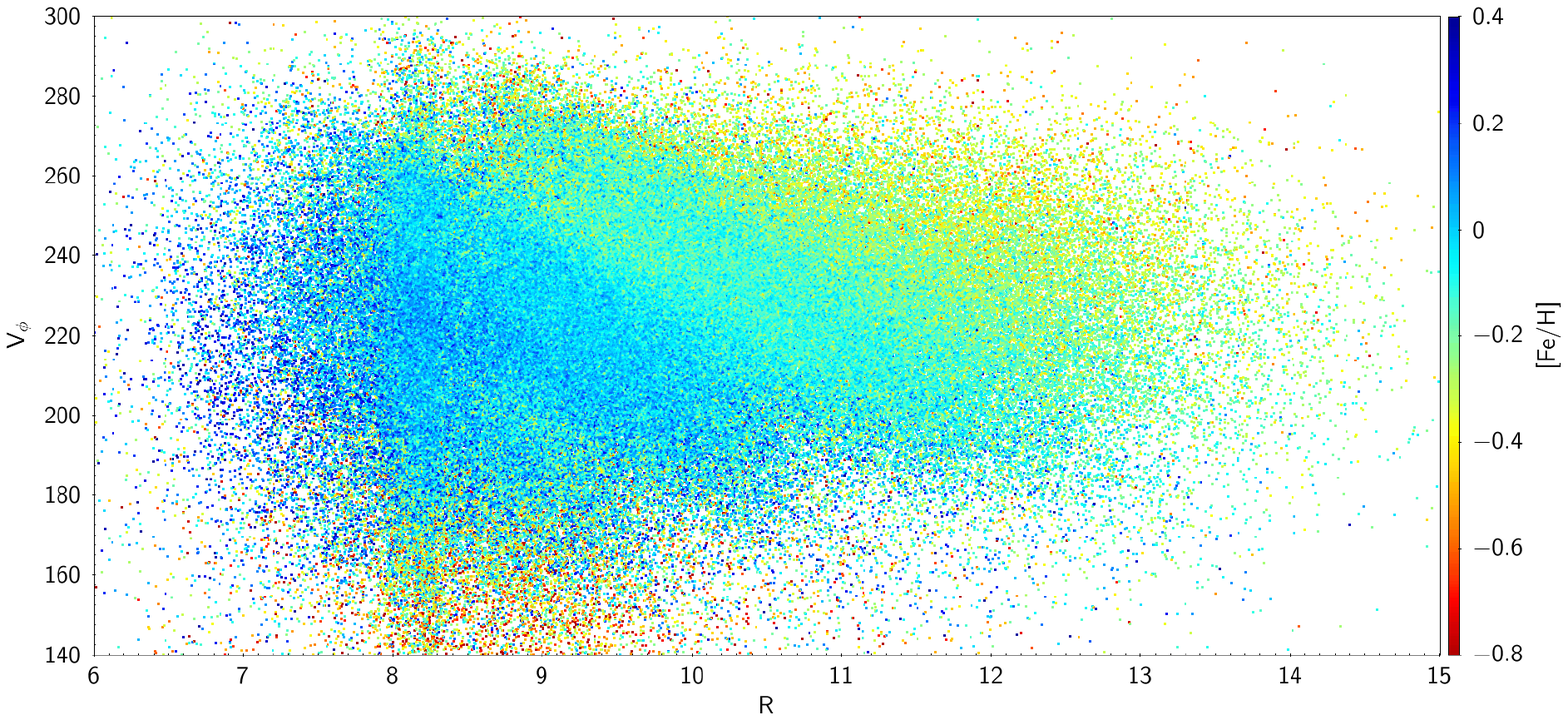}
\hspace{0pt}
\plotone{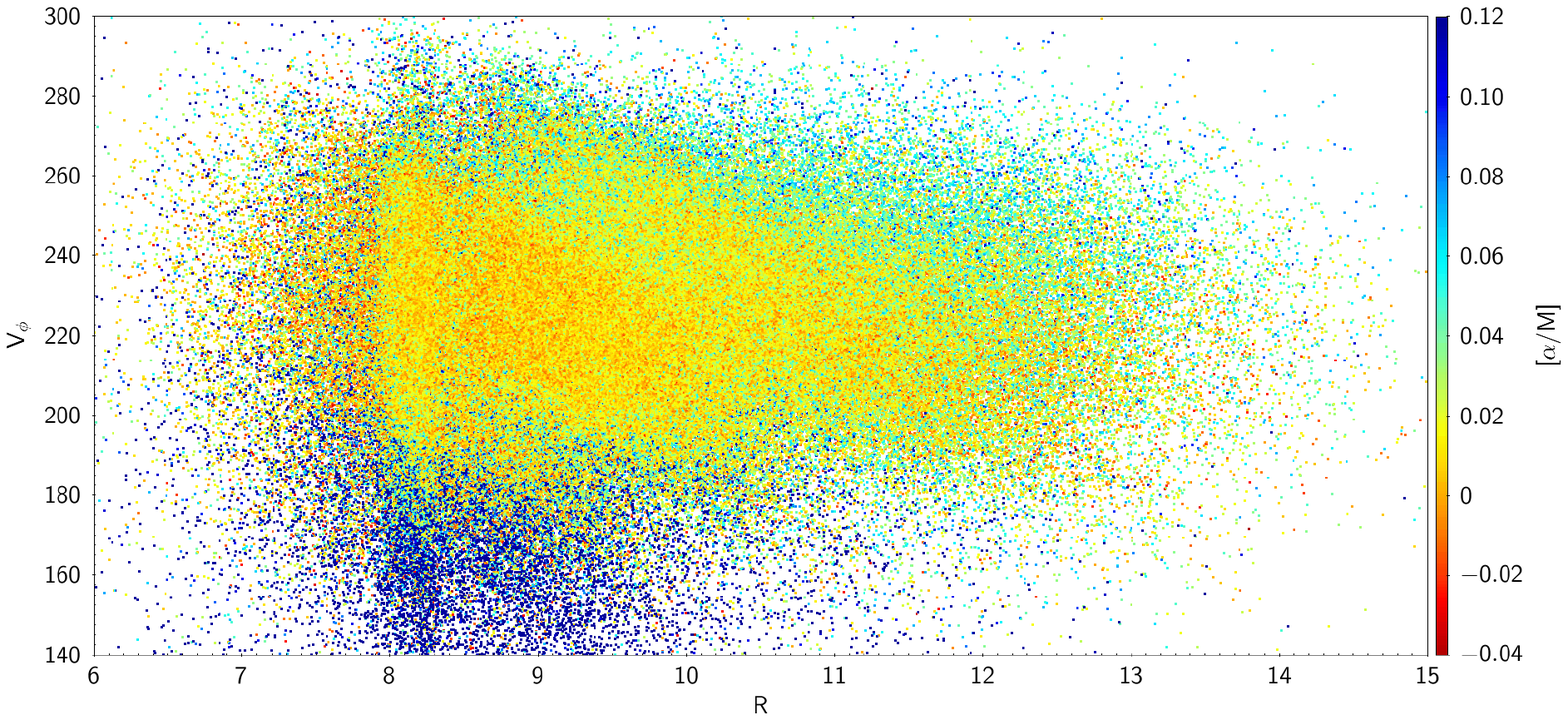}
\caption{$V_R$, [Fe/H] and [$\alpha$/M] distributions in the $V_{\phi}$ versus \textit{R} coordinates. As the color bars on the right show, different colors represent different median values of small bins.\label{fig:rvfifea}}
\end{figure}

\begin{figure}[htpb]
\begin{center}
\epsscale{.75}
\plottwo{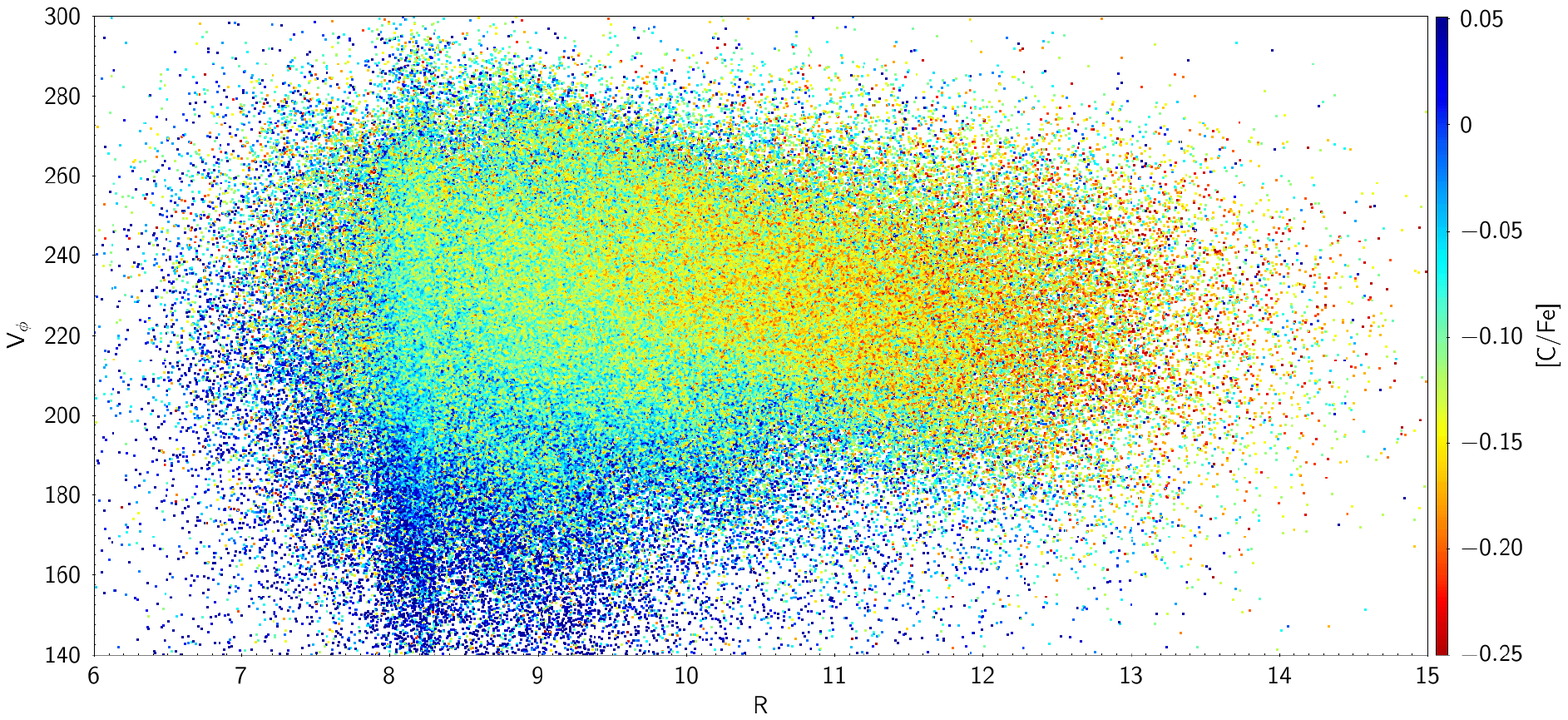}{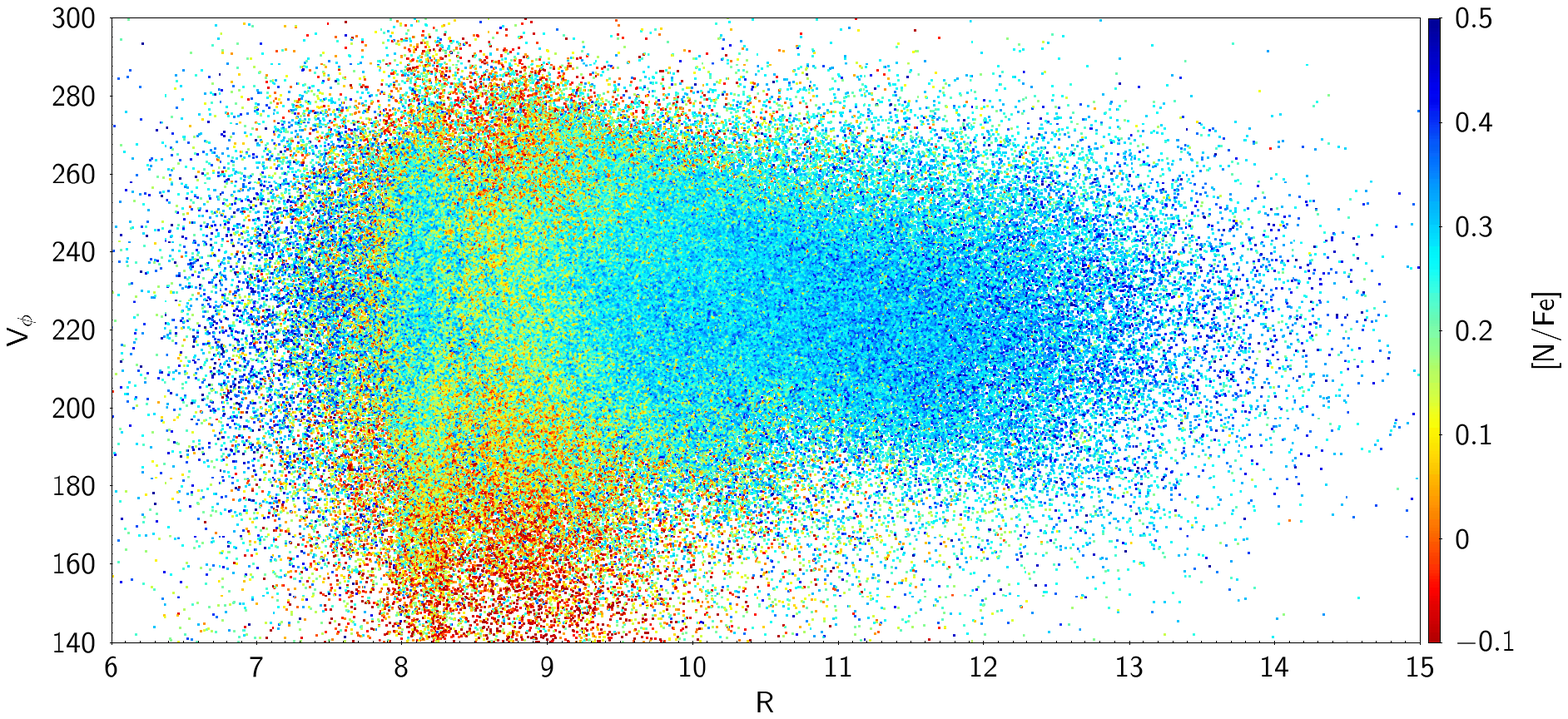}
\vspace{0pt}
\hspace{0pt}
\plottwo{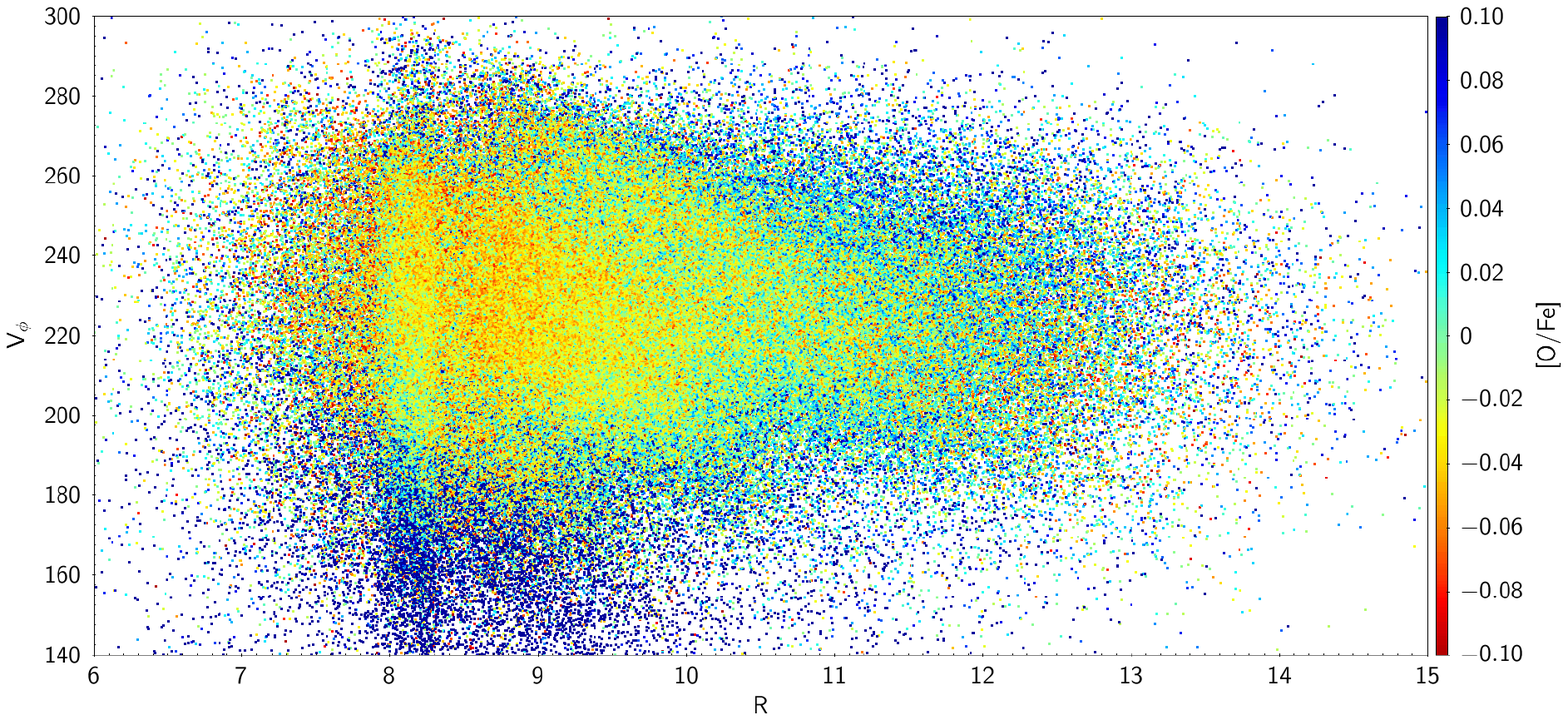}{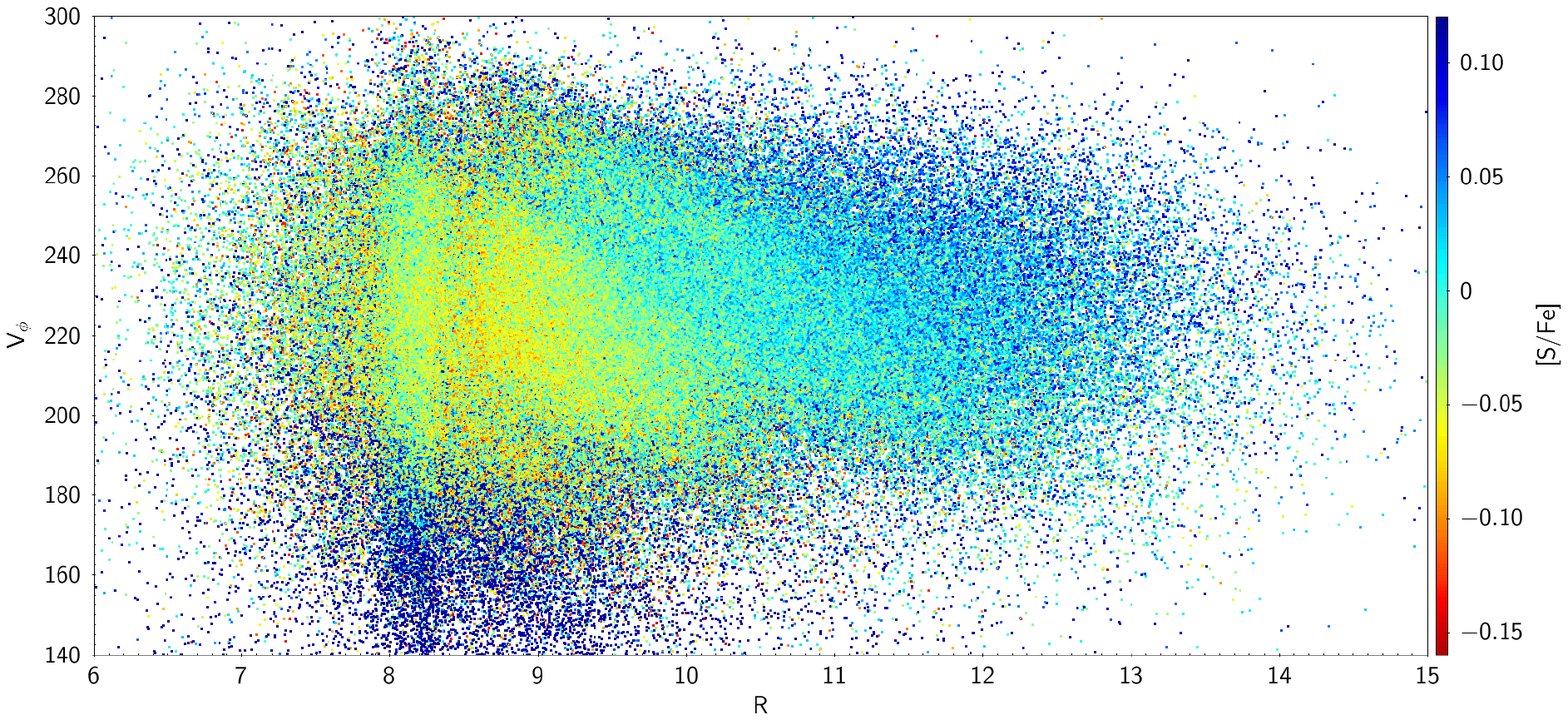}
\vspace{0pt}
\hspace{0pt}
\plottwo{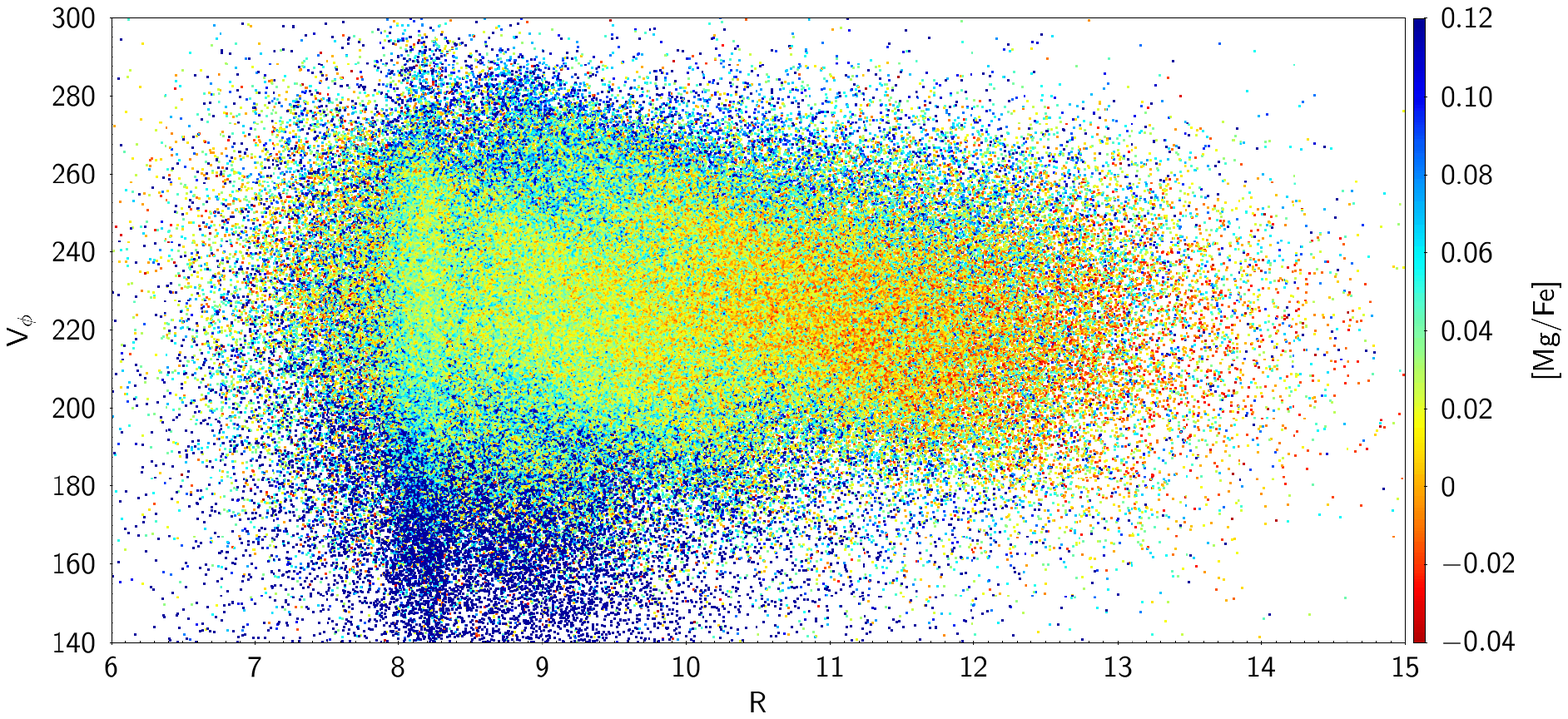}{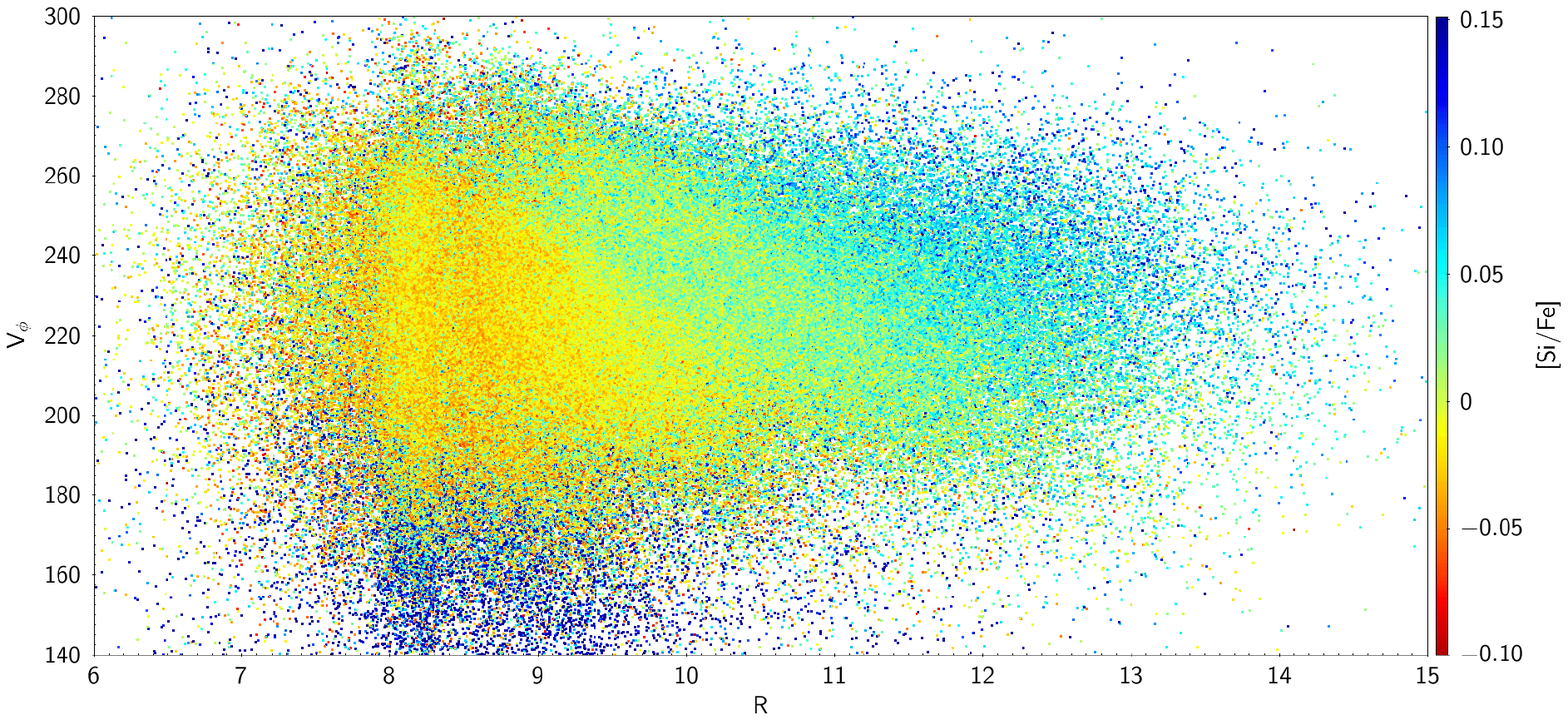}
\vspace{0pt}
\hspace{0pt}
\plottwo{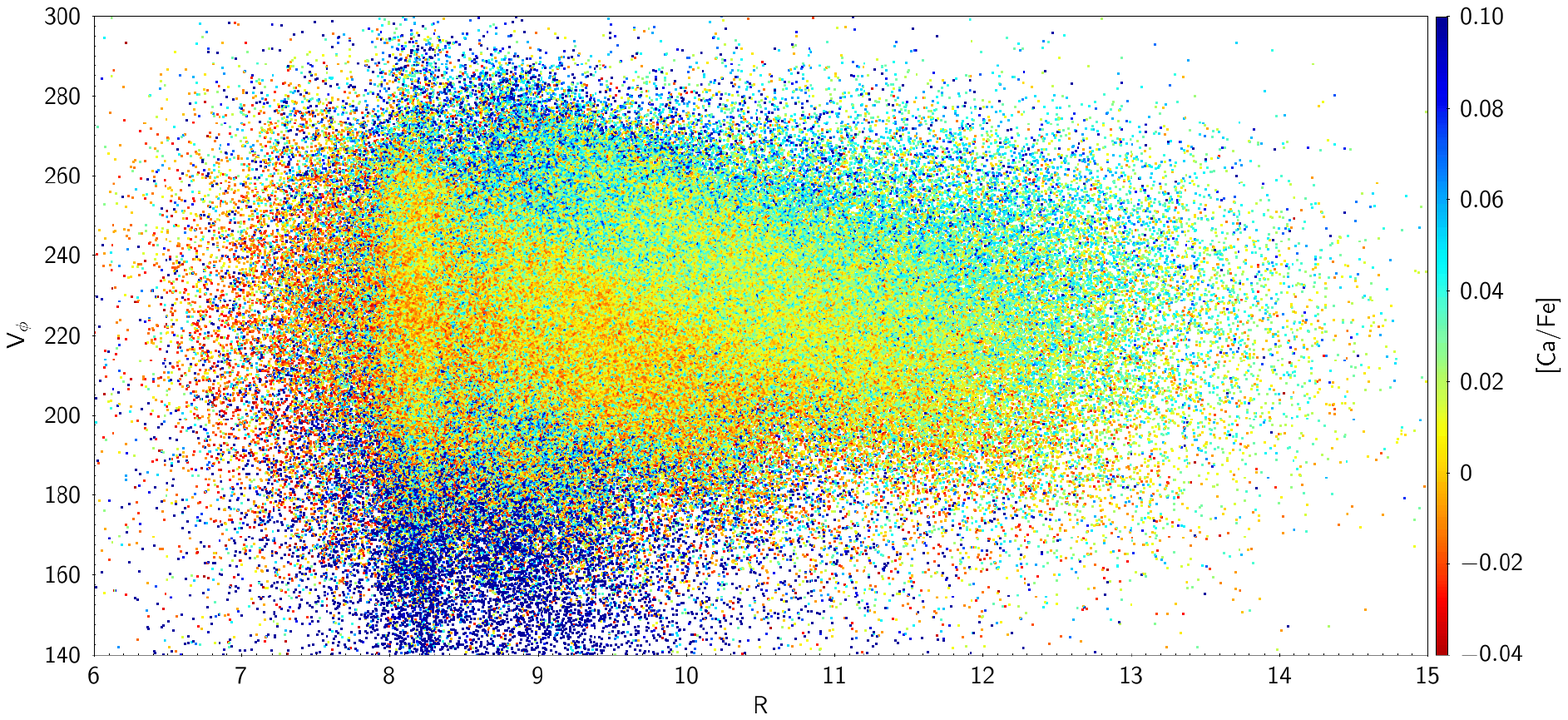}{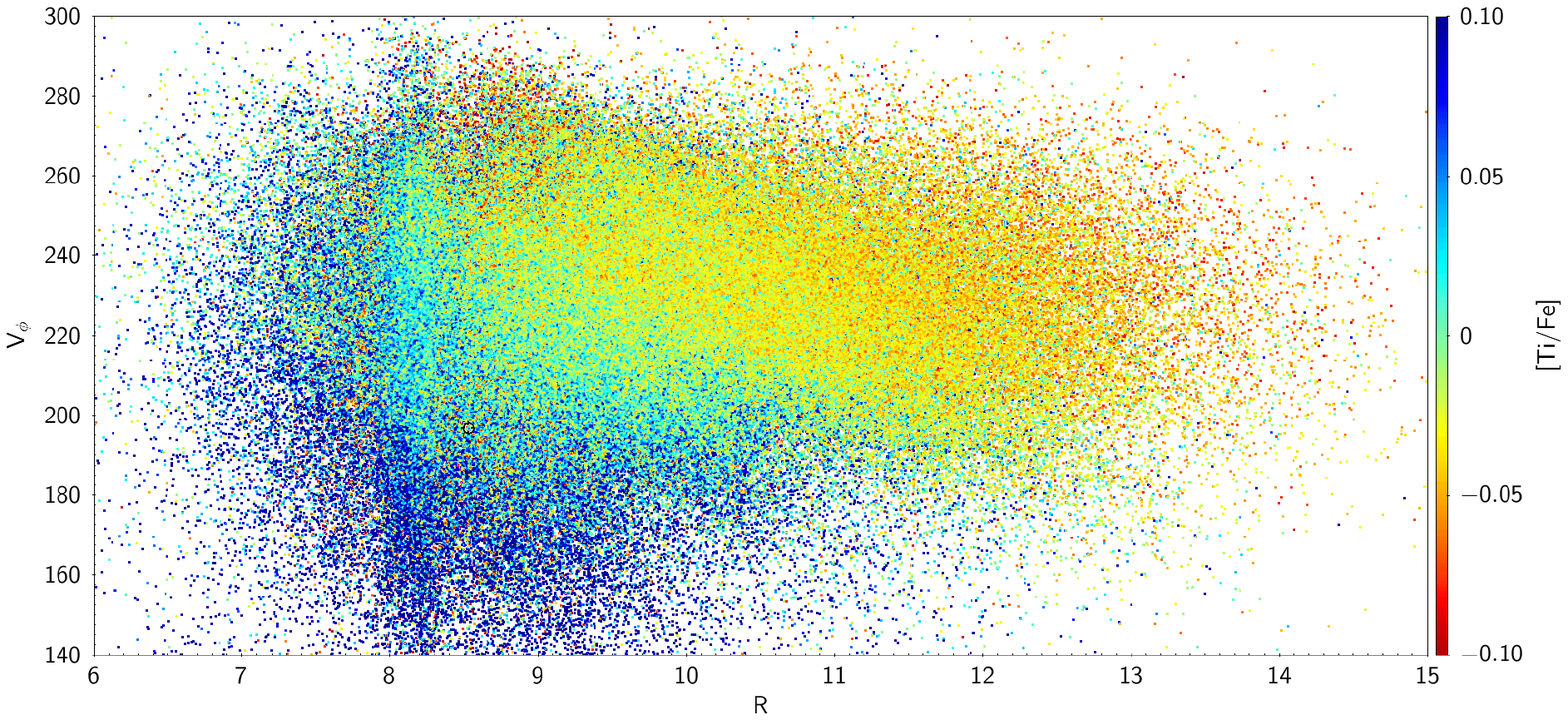}
\vspace{0pt}
\hspace{0pt}
\plottwo{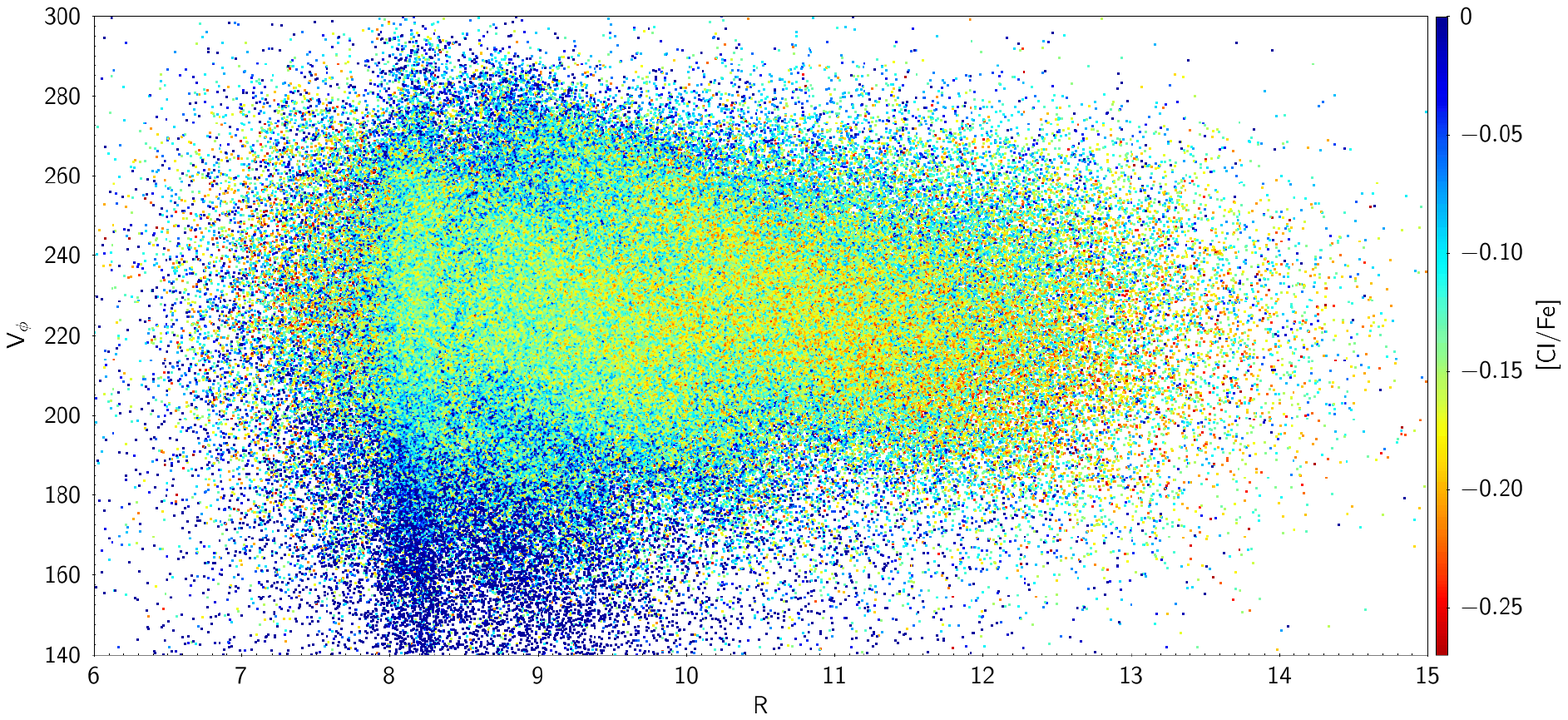}{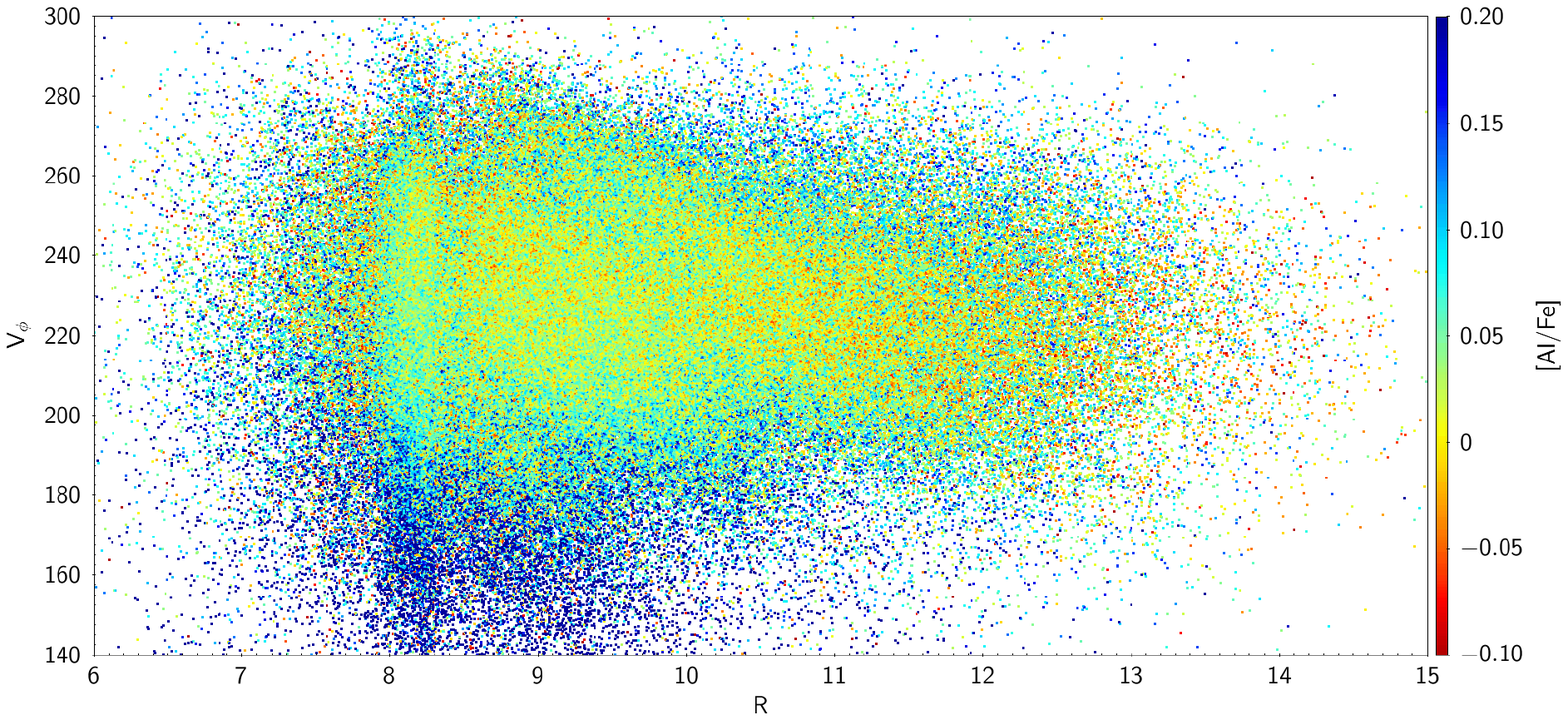}
\vspace{0pt}
\hspace{0pt}
\plottwo{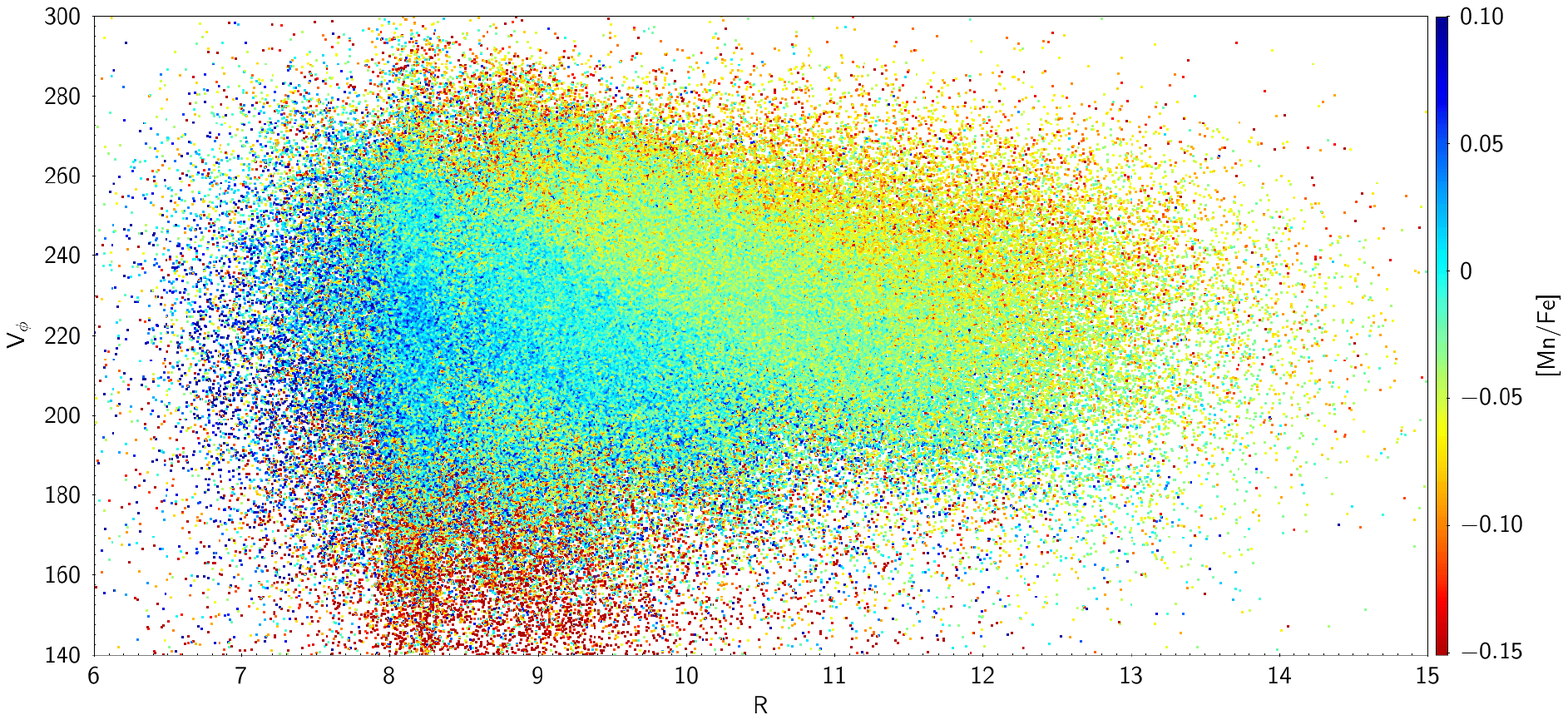}{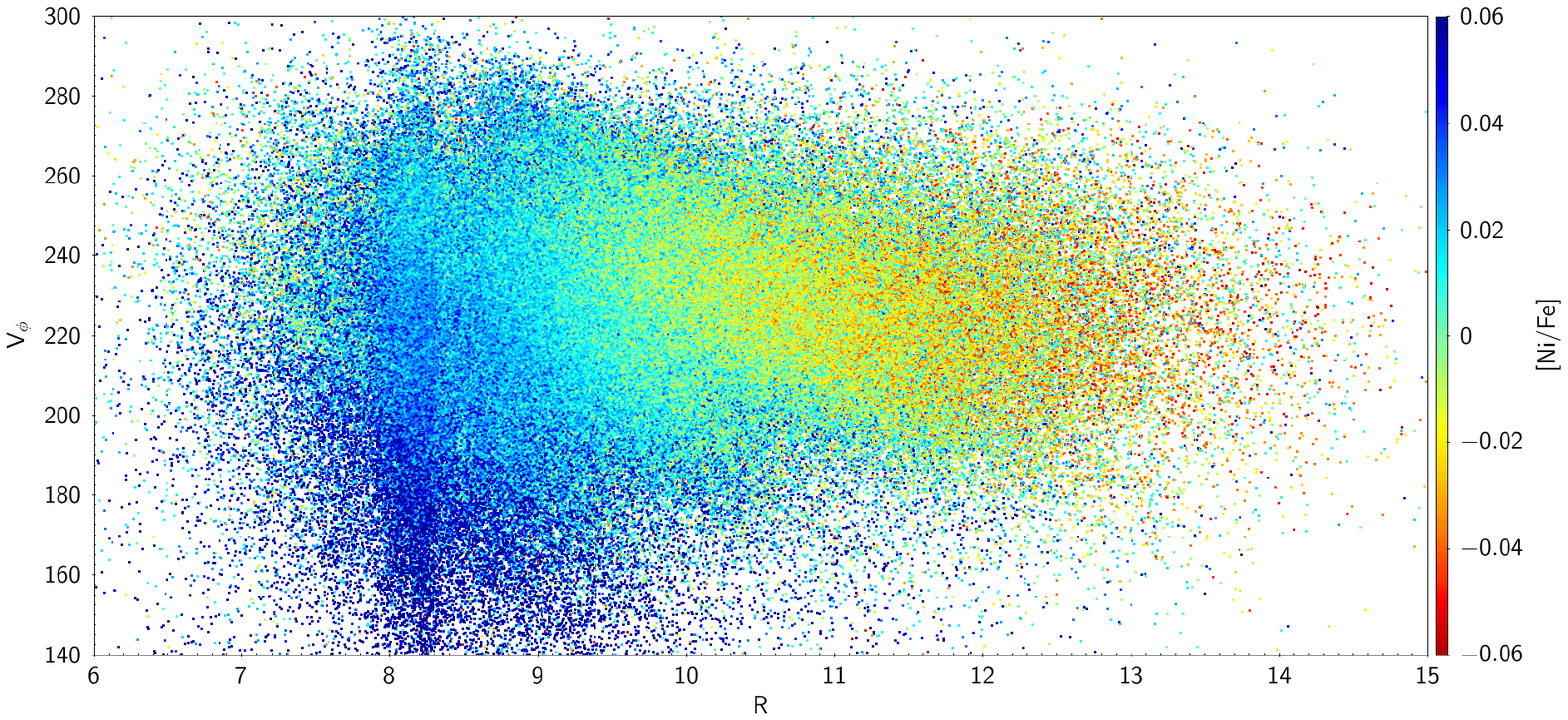}
\caption{Other elemental abundances distribution in the $V_{\phi}$ versus \textit{R} coordinates. \label{fig:rvfi}}
\end{center}
\end{figure}

Since Gaia DR2 been released, new small-scale and large-scale patterns in the phase space have been found and extensively studied. Kinematic and dynamic parameters have been used to reveal those structures in phase space. Chemical abundances can provide help to understand the nature of these features. \citet{kha19} used the GALAH survey data and found in (R, $V_{\phi}$) plane ridges can be clearly represented by [Fe/H] and [$\alpha$/Fe]. Figure \ref{fig:rvfifea} shows $V_R$, metallicity and $\alpha$ abundance distributions in the (R, $V_{\phi}$) plane and median values of weight parameters in small bins are color scaled by labels of color bars on the right of each subplot. The aim of taking medians to show elemental distributions is to alleviate the effect of spatial distribution of our target sample. TOPCAT \citep{tay05} software has been used to draw those subplots in figure \ref{fig:rvfifea} and  figure \ref{fig:rvfi} because it can automatically draw plots color scaled by median. Four different colors have been used to represent median $V_R$ value in each bin in the first subplot. As shown in the $V_R$ subplot, magenta and blue bins mean that more than half stars in those bins have $V_R > 0$, while orange and yellow bins mean that more than half stars in them have $V_R < 0$. Those black dash lines are hyperbolae that represent constant angular momentum tracks while white dash lines are diagonal lines. The signs of $V_R$ changes alternatively around those evenly spaced white dash lines like a wave. The discontinuous behaviour around $R = 8.34$ kpc is because the spatial distribution of our sample is not smooth or continuous around our sun. The other two subplots of figure \ref{fig:rvfifea} show [Fe/H] and [$\alpha$/M] distributions and they look similar to those of \citet{kha19} but with wider ranges. Like the $V_R$ subplot, [Fe/H] and [$\alpha$/M] abundances decrease and increase alternatively

Figure \ref{fig:rvfi} shows other elemental abundances distribution in the $V_{\phi}$ versus \textit{R} coordinates. Elemental abundances distributions of [O/Fe], [S/Fe], [Si/Fe], [Ca/Fe], [Al/Fe] and [Mn/Fe] show weak tendency that ridges with abundances decrease and increase alternatively. For other elements, though their distributions have diagonal features, their abundances monotonically change with \textit{R}. For $\alpha$ elemental abundances such as oxygen, sulfur, silicon and calcium we confirmed that ridges are related to $\alpha$ poor stars. As for magnesium and chlorine  abundances, their distributions are apparently affected by radially decreasing abundances. Titanium and nickel have smoothly changing distribution and it is difficult tell whether diagonal features are related to abundance rich stars or radially decreasing abundances distribution. The distribution of nitrogen is dominated by [N/Fe] increasing radially. $\alpha$ elements and light elements are more efficient to reflect short term events due to their born and spreading mechanism. Though the alternative changing features are not very strong in our elemental abundance subplots, we think they are related to those diagonal ridges in the first subplot. In a word, we have extends the chemical characterization of the ridges in the (R, $V_{\phi}$) plane up to R$=$13 kpc toward the anti-center farther than former research like \citet{kha19} reaching only R $=$ 9 kpc. Moreover we have more elemental distributions than their two elemental abundances distributions. Several mechanisms can induce diagonal ridges presented such as dynamical resonances \citep{fra19} or the phase mixing induced by a outside perturbation \citep{ant18} or transient spirals with/without outside perturbation \citep{kha19}.

\subsection{Elemental abundances gradient}

\begin{figure}[htpb]
\epsscale{.80}
\plotone{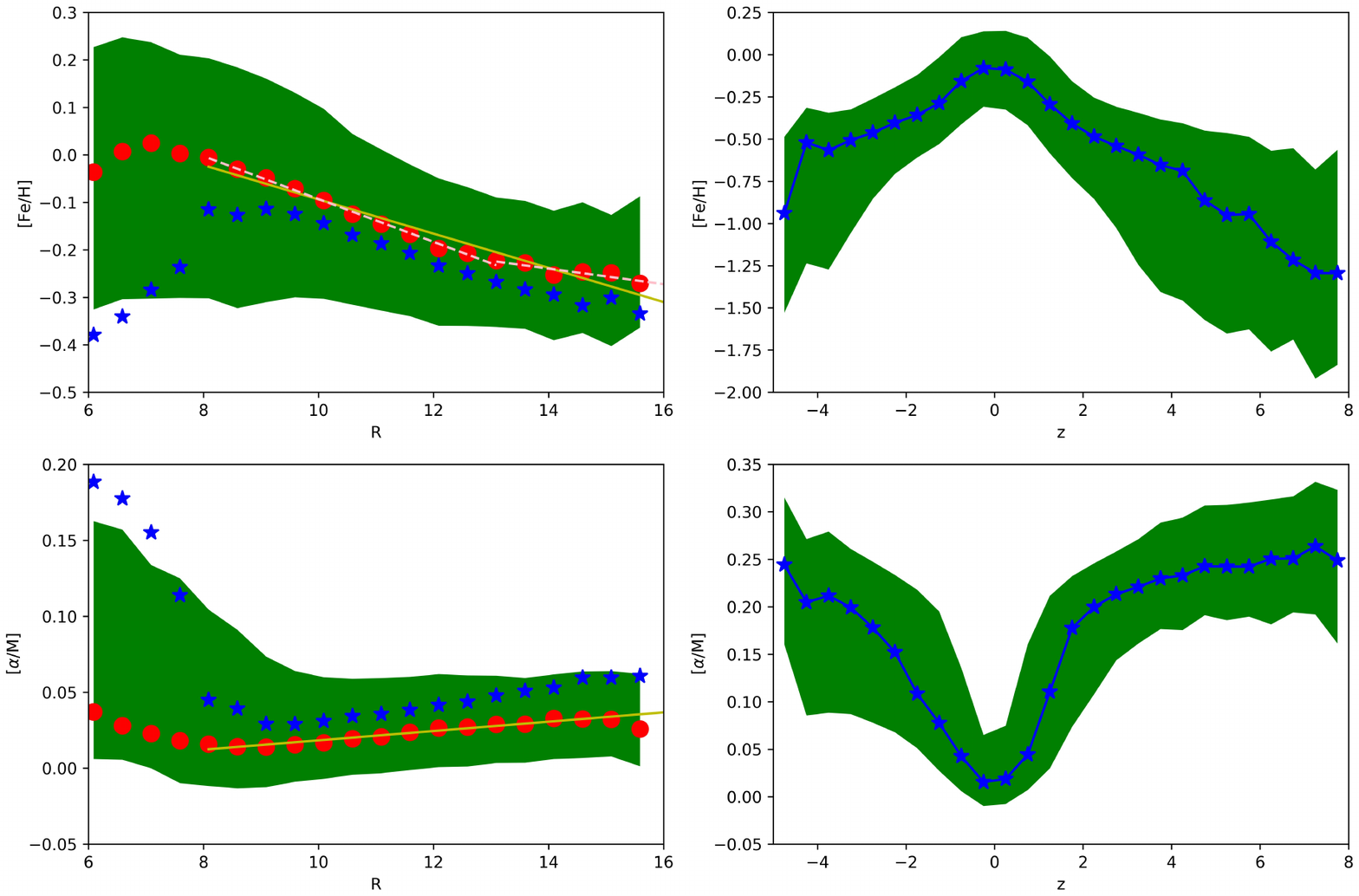}
\caption{Metallicity and $\alpha$ abundances distributions along R and z direction. Red dots represent medium values of disk stars in each bin and blue asterisks represent medium values of all stars of our sample in each bin. The green shaded areas show values between 16\% and 84\% quartiles of disk stars in the left column and of all stars in the right column.\label{fig:td1}}
\end{figure}

Figure \ref{fig:td1} shows metallicity and $\alpha$ abundances distributions along \textit{R} and \textit{z} directions. In the left column, bins with width $=$0.5 kpc are taken for stars with \textit{R} between (5.84, 15.84) kpc. In the right column, bins with width $=$0.5 kpc are taken for stars with \textit{z} between (-5, 8) kpc. Medium values of abundances of each bin are taken to represent the abundances of the position at the center of each bin. Blue asterisks in all subplots represent medium values of all stars of our sample in each bin while red dots represent medium values of disk stars with abs(\textit{z})$<$0.5 kpc in each bin. As farther way from galactic plane, stellar components changes from disk star dominated to halo star dominated. Therefore the elemental abundances distribution along z is complicated and we did not fit a line for them since there is no constant gradient. The green shaded areas show areas between 16\% and 84\% quartiles of disk stars in the left column and that of all stars in the right column. We only fitted lines for red points which look obviously linear related between elemental abundances and galactic radius. Abundance distribution with radius inner than the sun deviating from the linear relation is mainly caused by spatial distribution of LAMOST spectral survey. Those blue asterisks show spatial distributions of stellar abundances are clearly affected by other stellar components such as thick disk and halo. Thus for red dots inner than the sun, only the one next to the sun is used for linear fitting. In the upper left subplot of figure \ref{fig:td1}, the yellow line represents linear fit for red dots from R $=$ 8.09 kpc to 15.59 kpc while two pink dashed segments represent two segments fitting with Break-Point at 13.09 kpc. The fitted slope for yellow line is -0.0374 with standard error equal to 0.0020 and the slope is taken as radial gradient. The coefficient of determination given by lm function of R software\footnote{\url{ https://www.R-project.org}} is Multiple R-squared $= $0.9633 with p-value $= 1.035*\exp(-10)$. Segmented function from R software has been used to test two segments fitting for [Fe/H]. The suggested Break-Point by davies.test function is 13.09 with p-value $< 2.2*\exp(-16)$ and two slopes are respectively -0.0475 $\pm $0.0015 and -0.0173 $\pm $0.0028. Judging by p-value, two segments fitting is better than linear fitting for [Fe/H] versus \textit{R} which is consistent with \citep{yon12}. In the bottom left subplot, the last red dot is apparently deviated from the linear relation, so it is not used when we fit radial gradient for [$\alpha$/M]. The reason for the dot at 15.59 kpc deviates from the linear relation may be small quantity of stars in that bin and large uncertainties of parameters and we exclude it from linear fitting for other elements too. The radial gradient for [$\alpha$/M] is 0.0030 $\pm $0.0002 with Multiple R-squared $= $0.9519.

\begin{figure}[htpb]
\epsscale{.80}
\plottwo{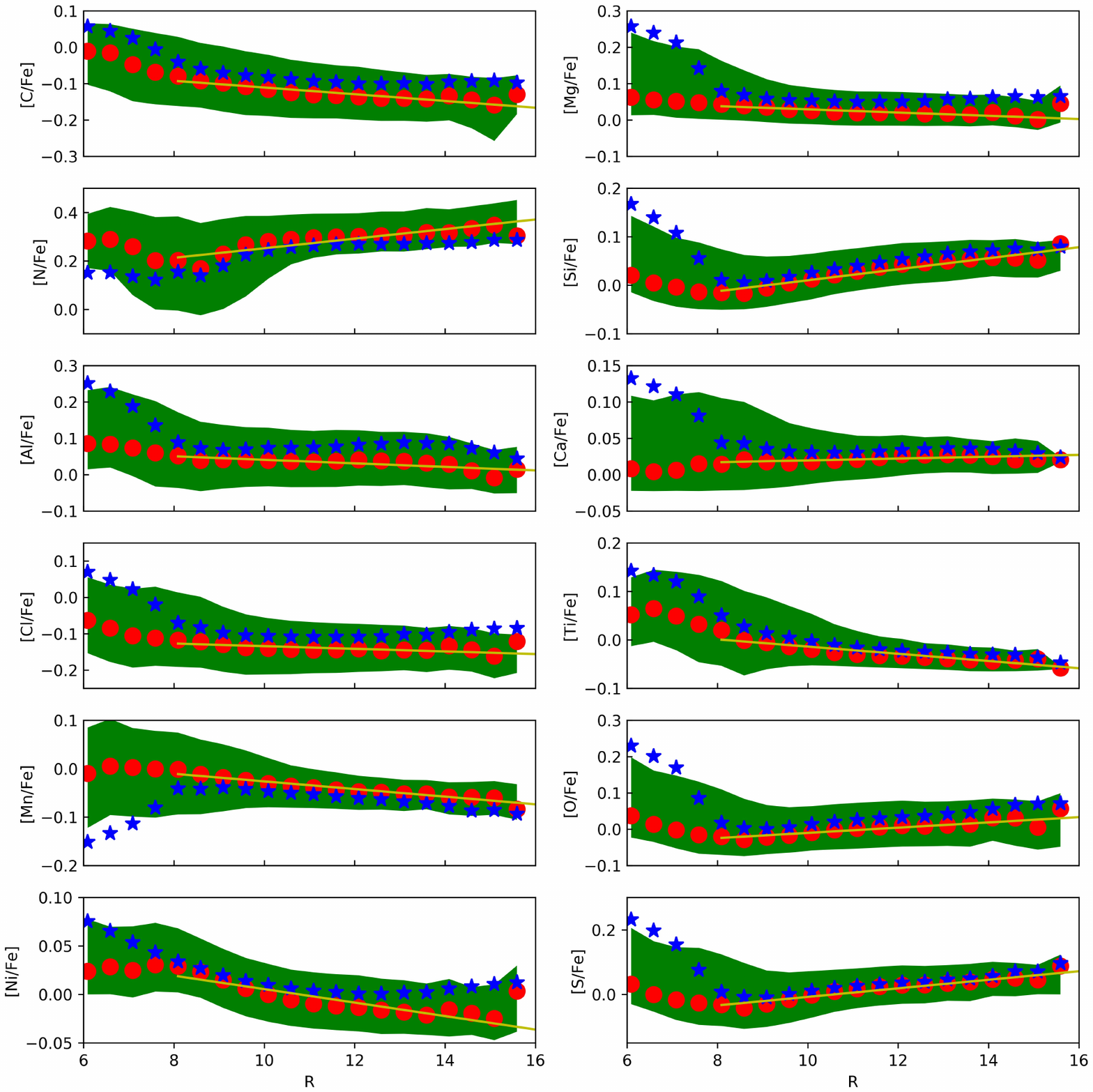}{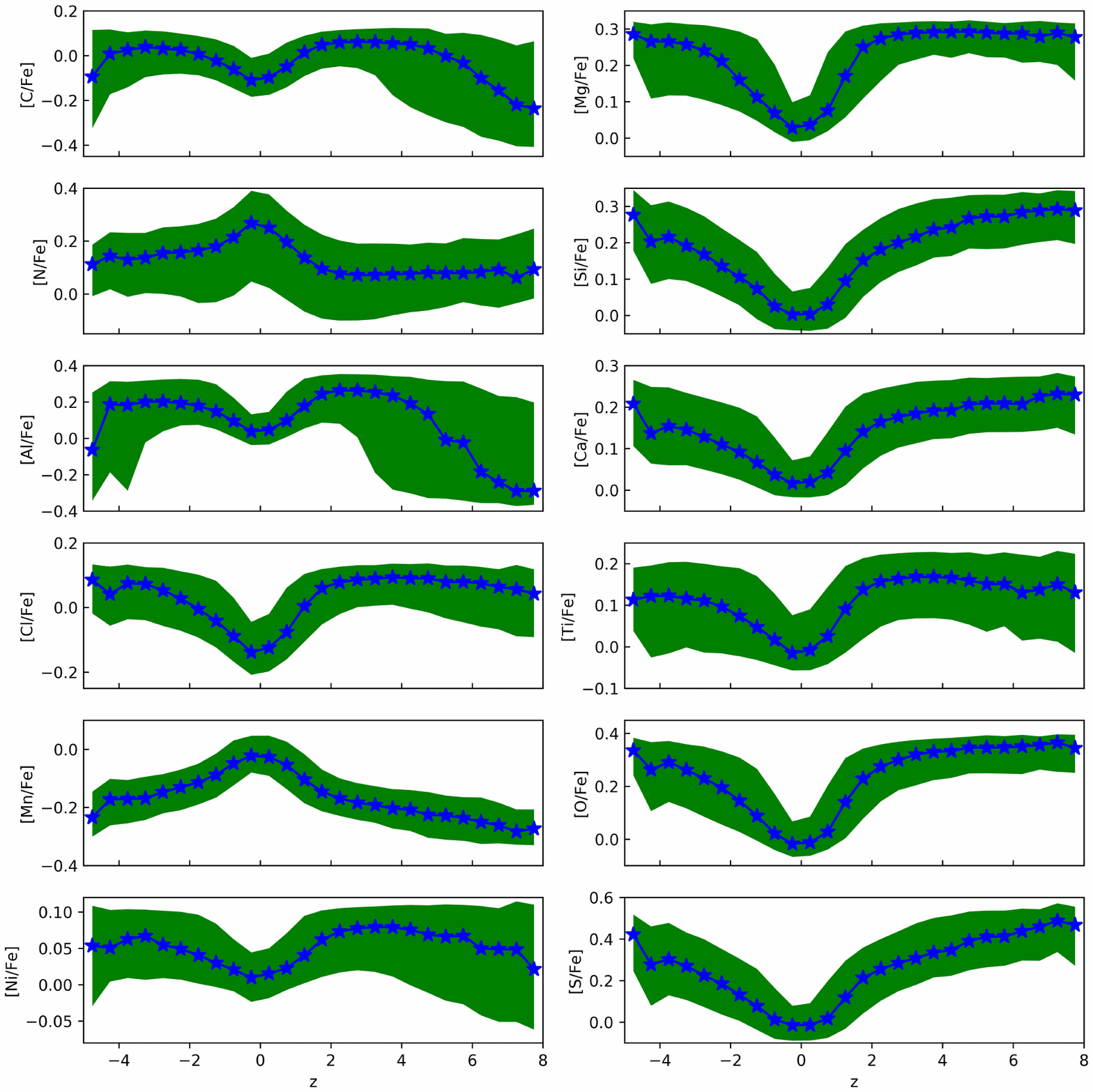}
\caption{Other elemental abundances distributions along R and z direction. Symbols are the same as in figure \ref{fig:td}.\label{fig:td}}
\end{figure}

Figure \ref{fig:td} shows other elemental abundances distributions along \textit{R} and \textit{z} direction. Our slopes for each fitted lines are respectively [C/Fe]: -0.0091 $\pm$ 0.0011, [N/Fe]: 0.0196 $\pm$ 0.0028, [O/Fe]: 0.0079 $\pm$ 0.0006, [Mg/Fe]: -0.0043 $\pm$ 0.0004, [Al/Fe]: -0.0039 $\pm$ 0.0008, [Si/Fe]:  0.0117 $\pm$ 0.0008, [S/Fe]: 0.0135 $\pm$ 0.0010, [Cl/Fe]: -0.0034 $\pm$ 0.0010, [Ca/Fe]: 0.0013 $\pm$ 0.0004, [Ti/Fe]: -0.0079 $\pm$ 0.0010, [Mn/Fe]: -0.0078 $\pm$ 0.0005, [Ni/Fe]: -0.0072 $\pm$ 0.0008. Most of them are very close to zero and some elemental abundances become richer radially others become poorer radially. Former researchers such as \citet{yon12} using open clusters and Cepheids with high resolution spectra to measure radial elemental abundances gradients which are more accurate than ours. Our advantage lies in that our sample shows continuously distribution along galactic radius. The whole tendency of our elemental abundances versus galactic radius is similar to former researches' results though gradient values are not the same. In conclusion, most of our predictions are usable and our data can effectively reveal statistical trends. As can be seen in the left two columns, distributions of elemental abundances versus \textit{z} are approximately symmetric with respect to $z = 0$. But they are not exactly symmetric so we think it is better to show elemental abundances versus \textit{z} than elemental abundances versus $\mid z \mid$. $\alpha$ elemental abundances become larger with distance departing from the galactic mid-plane. This reflects the stellar population changes along \textit{z} with stars of each bin changing from disk dominated to halo dominated. Therefore it is complicated for abundances versus \textit{z} and we do not fit lines for them.

\section{CONCLUSION}

We used an open source code called astroNN to train neural networks for common stellar sample of APOGEE DR14 and LAMOST DR5. Then with trained neural networks, 12 elemental abundances have been predicted for LAMOST DR5 spectra of our target sample. An over-sampling technique has been used to generate training and test set for each element from common stellar sample. We have shown our training and test results analysis and for most stars, residuals are small. For predictions with large residuals, their possible reasons have been analysed that mostly related to nonuniformity of common stellar sample. Nonetheless, uncertainties are provided to be used to select reliable predictions. After obtained elemental abundances for LAMOST spectra, we cross matched them with Gaia DR2 catalog to acquire full space and velocity information. Then the thin disk stars were analyzed chemically in $V_{\phi}$ versus \textit{R} coordinate. We found chemical abundances distributions in the (R, $V_{\phi}$) plane may be related to kinematic ridges up to large galactocentric distances. With our data, it is still very difficult to reach a final conclusion on which mechanisms cause the ridges. Several mechanisms could explain the results presented here such as classical dynamical resonances of the bar, spiral perturbation or the phase mixing induced by a perturber. We have not done any n-body simulations but our results can help to interpret future new detailed simulations which should include chemical evolutions in them.

Finally, we have shown elemental abundances distributions along R and z direction. Only radial elemental abundances gradients have been calculated for disk stars because elemental abundances relative to \textit{z} are complicated.

\acknowledgments
We thank Bharat Kumar Yerra, JiYeon Seok and Kefeng Tan for their suggestions. This study is supported by the National Natural Science Foundation of China under grant No. 11988101, 11973048, 11927804, 11890694, 11573035, 11625313 and National Key R\&D Program of China No. 2019YFA0405502. This work is also supported by the Astronomical Big Data Joint Research Center, co-founded by the National Astronomical Observatories, Chinese Academy of Sciences and the Alibaba Cloud. This project was developed in part at the 2018 Gaia-LAMOST Sprint workshop, supported by the National Natural Science Foundation of China (NSFC) under grants 11333003 and 11390372. Guoshoujing Telescope (the Large Sky Area Multi-Object Fiber Spectroscopic Telescope LAMOST) is a National Major Scientific Project built by the Chinese Academy of Sciences. Funding for the project has been provided by the National Development and Reform Commission. LAMOST is operated and managed by the National Astronomical Observatories, Chinese Academy of Sciences.

\end{document}